\documentclass[a4paper,fleqn,usenatbib]{mnras}

\usepackage{newtxtext,newtxmath}

\usepackage[T1]{fontenc}
\usepackage{ae,aecompl}

\usepackage{graphicx}	
\usepackage{amsmath}	
\usepackage{enumitem}
\usepackage{float}

\graphicspath{{NSpics_2/}}

\title[Neutron Star Mountain Creation]{Mountain formation by repeated, inhomogeneous crustal failure in a neutron star}

\author[A. D. Kerin et al.]{
A. D. Kerin,$^{1}$\thanks{E-mail: akerin@student.unimelb.edu.au}
A. Melatos$^{1,2}$
\\
$^{1}$School of Physics, University of Melbourne, Parkville, VIC 3010, Australia \\
$^{2}$Australian Research Council Centre of Excellence for Gravitational Wave Discovery (OzGrav), \\University of Melbourne, Parkville, VIC 3010, Australia}
\date{Accepted 2022 May 10. Received 2022 May 9; in original form 2021 March 28}

\pubyear{2022}

\begin{document}
\label{firstpage}
\pagerange{\pageref{firstpage}--\pageref{lastpage}}
\maketitle

\begin{abstract}
The elastic crust of a neutron star fractures repeatedly as it spins down electromagnetically. An idealised, macroscopic model of inhomogeneous crustal failure is presented based on a cellular automaton with nearest-neighbour tectonic interactions involving strain redistribution and thermal dissipation. Predictions are made of the size and waiting-time distributions of failure events, as well as the rate of failure as the star spins down. The last failure event typically occurs when the star spins down to $\approx 1\%$ of its birth frequency with implications for rotational glitch activity. Neutron stars are commonly suggested as sources of continuous gravitational waves. The output of the automaton is converted into predictions of the star's mass ellipticity and gravitational wave strain as functions of its age, with implications for future observations with instruments such as the Laser Interferometer Gravitational Wave Observatory (LIGO), the Virgo interferometer, or the Kamioka Gravitational Wave Detector (KAGRA).
\end{abstract}




\section{Introduction}
\label{sec:Introduction}

Mountains form on neutron stars through various thermoelastic and magnetic forces \citep{ushomirsky2000deformations, payne2004burial, haskell2008modelling, mastrano2012updated, gittins2021modelling}. They break the star’s axisymmetry and lead to the continuous emission of gravitational waves at levels potentially approaching the sensitivity of instruments like the Laser Interferometer Gravitational Wave Observatory (LIGO), the Virgo interferometer or the Kamioka Gravitational Wave Detector (KAGRA). \citep{riles2013gravitational, woan2018evidence, reed2021modeling}. Several recent LIGO-Virgo-KAGRA searches have placed increasingly stringent constraints on the mass ellipticity of neutron star mountains \citep{abbott2019narrow, abbott2019directional, abbott2019search,abbott2019searches1,  abbott2019searches2, abbott2020gravitational, papa2020search}, and more sensitive searches are planned in the near future. More broadly, mechanical failure in response to crustal deformation \citep{baym1971neutron, pines1972microquakes} has been invoked in the context of numerous transient astrophysical phenomena such as rotational glitches in pulsars \citep{ruderman1976crust, middleditch2006predicting}, soft gamma-ray repeaters \citep{kaplan2001hubble, hurley2005exceptionally, baiko2018breaking} and fast radio bursts \citep{suvorov2019young}. The connection between crustal deformation, mechanical failure, and the above phenomena remains a matter of debate. Viable alternatives exist, e.g. superfluid vortex avalanches in the context of pulsar glitches \citep{anderson1975pulsar, warszawski2011gross}. Predictive, falsifiable models are needed to discriminate between the alternatives, based either on the microphysics \citep{andersson2012pulsar, piekarewicz2014pulsar} or on the long-term size and waiting-time statistics of the transients \citep{fulgenzi2017radio,melatos2018size,carlin2020long}.
 

The far-from-equilibrium \textit{tectonic} process by which mountains form on macroscopic scales is currently unknown \citep{caplan2017colloquium, gittins2021modelling}. To date modelling has concentrated on calculating the microphysics of crust failure \citep{horowitz2008molecular, horowitz2009breaking, chugunov2010breaking, horowitz2011neutron, baiko2018breaking}. Molecular dynamics simulations of $\sim 10^7$ ions responding to a slowly ramping, quasistatic stress indicate that the crust fails globally (above a strain of $\approx 0.1$) rather than fracturing locally into cracks due to the high pressure \citep{horowitz2009breaking}. Some modelling has treated the whole star on macroscopic scales as it evolves through a quasistatic sequence of elastic and hydromagnetic equilibria. \citet{giliberti2019incompressible} and \citet{giliberti2020modelling} studied two-layer and continuously stratified models respectively and evaluated the effect of the adiabatic index and stellar mass on the build up of mechanical strain in the crust. However, the latter papers describe slow, reversible evolution on the spin-down time-scale. They do not include the stick-slip dynamics of repeated failure and healing, which involve a mixture of short and long time-scales and lead to irreversible, history-dependent outcomes, i.e. hysteresis \citep{jensen1998self, sornette2006critical}.

In this paper we address the problem of mountain formation by repeated crustal failure on macroscopic scales driven by spin down. The paper is structured in the following way. Section \ref{sec:Tectonics} deals with how the crust deforms secularly due to spin down and the redistribution of stress and dissipation of energy in response to failure, which leads to radial movement. A cellular automaton is constructed to capture the dynamics of repeated failure. Section \ref{sec:Failureeventstatisitcs} presents the long-term statistics of failure events generated by the automaton. Section \ref{sec:MassEllipticity} calculates the ellipticity and hence gravitational wave strain as a function of the star's age. The calculations in this paper do not include magnetic fields for simplicity. Complementary studies of plate tectonics involving the interplay between superfluid neutron vortices and magnetic flux tubes are presented by \citet{ruderman1991neutron1, ruderman1991neutron2, ruderman1991neutron3}; see also \citet{srinivasan1990novel}.

\section{Tectonics}
\label{sec:Tectonics}

A neutron star loses angular momentum over its lifetime due to processes such as magnetic dipole radiation and gravitational wave emission \citep{lyne201445}. As the balance of gravitational and centrifugal forces changes, so does the shape of the star \citep{pines1972microquakes}. A neutron star is primarily composed of an incompressible superfluid interior and a solid crystalline crust \citep{horowitz2011neutron, fattoyev2018crust, baiko2018breaking, chugunov2020neutron, kozhberov2020deformed, kozhberov2020breaking}. The superfluid interior adjusts to spin-down deformation, but the crust, being an almost-rigid solid, does not deform as easily. Mechanical strain builds up in the crust up to and beyond the point of failure.

We propose that the accumulated strain causes sections of the crust to fail when the breaking strain is exceeded locally \citep{franco2000quaking, fattoyev2018crust}. The failed sections move radially, creating a mountain locally as well as dissipating energy and redistributing strain to neighbouring sections, which may then fail themselves in response. These knock-on dynamics are modelled by a cellular automaton described in Section \ref{sec:CellAutomaton}.

\subsection{Secular crustal deformation without failure}
\label{sec:SecDefNoFail}

In the absence of failure, we consider a biaxial star, which is axisymmetric and rotating about the $z$-axis. The star is composed of a fluid core and solid, almost-rigid crust, whose principal moments of inertia vary quasistatically on the spin-down time-scale, which is much longer than the rotation period relevant in gravitational wave
applications, for example.

The slow, secular deformation of the star is calculated using the formalism outlined by \cite{franco2000quaking}. Briefly, this approach starts by equating the sum of the gravitational and centrifugal forces to the elastic forces and considers a Lagrangian perturbation to that equilibrium. To summarise the main result, when the star spins down an infinitesimal amount, $\delta\Omega$, from some angular velocity, $\Omega_{\rm i}$, to a smaller value, $\Omega_{\rm f}=\Omega_{\rm i}-\delta\Omega$, an infinitesimal volume element at $\vec{r}$ moves to $\vec{r}+\vec{u}(\vec{r})$, where $\vec{u}(\vec{r})$ is the deformation vector. Its components are given below in spherical coordinates,

\begin{eqnarray}
u_{r}(\vec{r})&=& \left( a r-\frac{A r^3}{7}-\frac{B}{2 r^2}+\frac{b}{r^4} \right)P_{2}(\cos \theta),\label{eq:defvecs1}\\
u_{\theta}(\vec{r})&=& \left( \frac{a r}{2}-\frac{5 A r^3}{42}-\frac{b}{3r^4} \right)\frac{dP_{2}(\cos \theta)}{d\theta},\label{eq:defvecs2}\\
u_{\phi}(\vec{r})&=&0,
\label{eq:defvecs3}
\end{eqnarray}
where $P_{2}(\cos \theta)=(3\cos^2\theta-1)/2$ is the second Legendre polynomial with argument $\cos\theta$. The coefficients $a, b, A, B$ satisfy the simultaneous equations,
\begin{eqnarray}
0&=&a-\frac{8AR^2}{21}-\frac{B}{2R^3}+\frac{8b}{3R^5},\label{eq:defCoeffs1}\\
0&=&a-\frac{8AR'^2}{21}-\frac{B}{2R'^3}+\frac{8b}{3R'^5},\\
0&=&-2f'(R)-\frac{2 V_{K}^2}{5 C^2 }\frac{f(R)}{R}+\frac{R^2 (\Omega_{\rm i}^2-\Omega_{\rm f}^2)}{3 C^2}
-AR^2-\frac{B}{R^3},\label{eq:defCoeffs3}\quad\\
0&=&-\frac{1}{2}\left( AR'^2+\frac{B}{R'^3} \right) -f'(R').
\label{eq:defCoeffs4}
\end{eqnarray} 
\\
In Eqs. (\ref{eq:defCoeffs1})--(\ref{eq:defCoeffs4}), $R$ and $R'<R$ are the total and core radii of the non-rotating configuration respectively. We also define $V_{\rm K} =\sqrt{GM/R}$ as the surface Keplerian velocity, and $C = \sqrt{\mu/\rho_{\rm crust}}$ as the transverse speed of sound, where $\mu$ is the shear modulus, $\rho_{\rm crust}$ is the crustal density, and $f(r)$ is simply the part of Eq. (\ref{eq:defvecs1}) that depends on $r$; $f'(r)$ is its derivative. In this paper we set $R = 10.5$ km, $R' = 9.5$ km, $V_{\rm K} = 1.4 \times 10^5$ kms\textsuperscript{-1},  $C = 1.55 \times 10^3$ kms\textsuperscript{-1} and assume a 1.4 solar mass star \citep{franco2000quaking, fattoyev2018crust}. These are illustrative values for intrinsically uncertain quantities; the goal of the paper is to introduce the idea of tectonic activity modelled by a cellular automaton, not to make precise quantitative predictions. The boundary conditions, Eqs. (\ref{eq:defCoeffs1})-(\ref{eq:defCoeffs4}), come from requiring that there are no shear forces at the crust-core and crust-vacuum boundaries and that the star is in mechanical equilibrium at those same boundaries.

The strain tensor,
\begin{eqnarray}
\alpha_{jk}(\vec{r})=\frac{1}{2}\left[ \frac{\partial u_{j}(\vec{r})}{\partial x_{k}}+\frac{\partial u_{k}(\vec{r})}{\partial x_{j}}\right],
\label{eq:Tensor}
\end{eqnarray}
defines the strain angle, $\xi(\vec{r})$, at every point in the crust as the difference between its largest and smallest eigenvalues. \cite{TheoryOfElasticity} discussed in detail the connection between $\alpha_{jk}(\vec{r})$ and $\xi(\vec{r})$. The latter quantity, $\xi(\vec{r})$, is the \textit{increase} in the total strain angle, $\gamma(\vec{r})$, at a particular location as a result of spinning down infinitesimally from $\Omega$ to $\Omega - \delta\Omega$, not the \textit{current} strain angle that has built up over the entire course of spin down, i.e. from $\Omega(t = 0)$ to $\Omega(t)-\delta\Omega$. 
In this paper we mostly refer to the strain angle (fractional deformation), rather than stress (force per unit area). However the two are linearly related up until the point of failure \citep{horowitz2008molecular,horowitz2009breaking}.

\begin{figure}
\includegraphics[height=5.5cm,width=8.5cm]{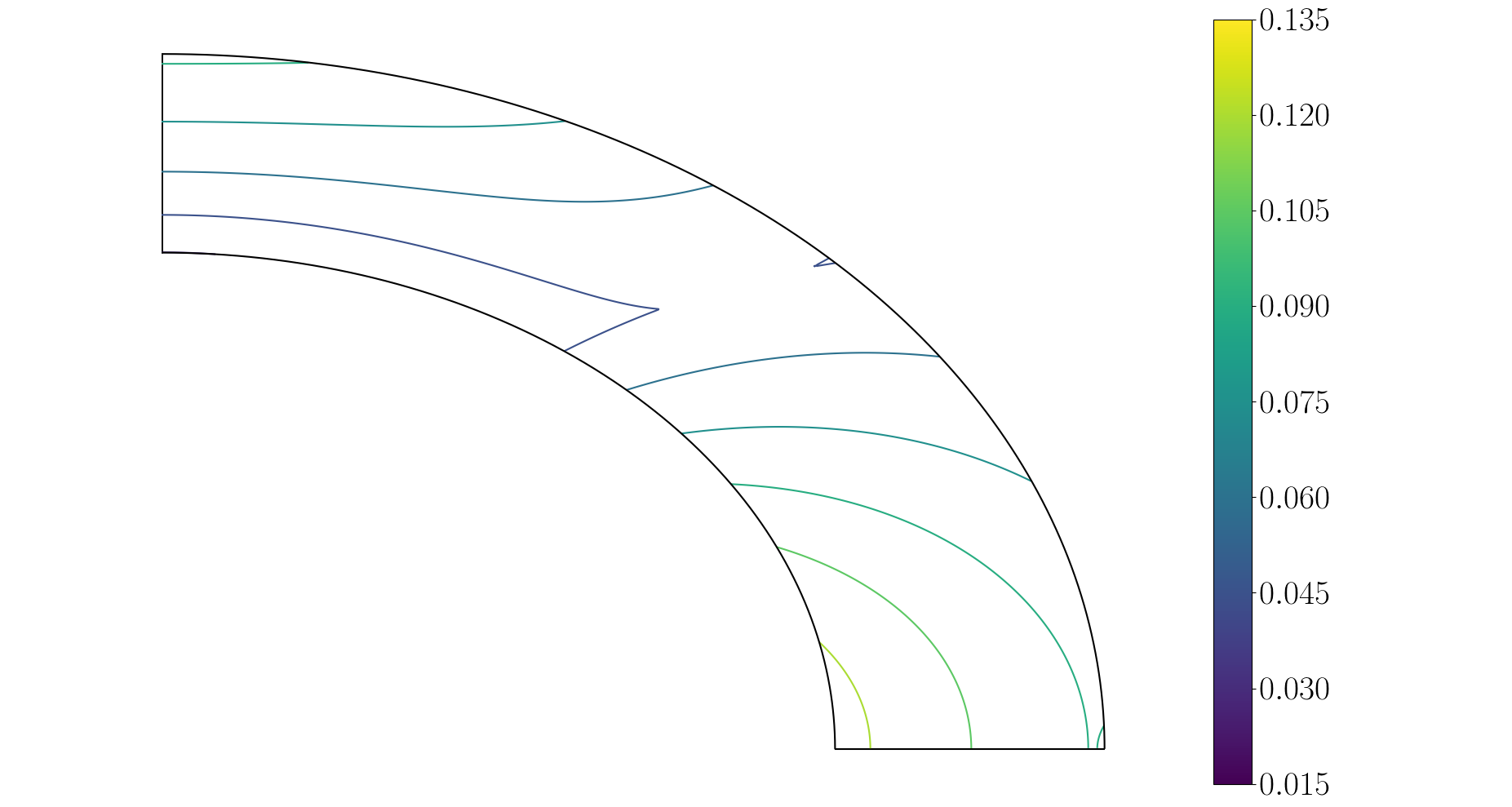}
\caption{
Contours of constant strain angle in a meridional cross-section of the crust from the north pole to the equator, evaluated using Eqs. (\ref{eq:defvecs1})-(\ref{eq:defCoeffs4}). The strain angle is greatest near the base of the crust at the equator. It is smallest near the base of the crust near the pole as well as throughout the crust at intermediate latitudes. In other (e.g. stratified) models the location of greatest strain may be different. The thickness of the crust has been exaggerated for legibility. The plotted contours indicate the strain angle that accumulates over the star's entire life i.e. from $\Omega(0) = 800$Hz to $\Omega(t \rightarrow \infty) = 0$.
}
\label{fig:ContourPlot}
\end{figure}

In Fig. \ref{fig:ContourPlot} we plot contours of the strain angle in the crust based on the model of \citet{franco2000quaking}. It is apparent that the strain angle is highest at the crust-core boundary at the equator. This property influences how the strain angle is evaluated in the automaton in Section \ref{sec:CellAutomaton}. In this model $\mu$ is constant throughout the crust for simplicity, but there are other models where this is not the case. For example in the papers by \citet{giliberti2019incompressible} and \citet{giliberti2020modelling}, which include mass stratification, the maximum strain occurs at the poles, the base of the crust, or the stellar surface, depending on the exact model. Likewise \citet{cutler2003crustal} found that the stress is greatest at the crust-core boundary yet the strain is greatest at the stellar surface, because $\mu$ increases with depth.

\subsection{Microphysics of failure of the crystal lattice}
\label{sec:Micro}
Simulations of the crustal material have been performed in the past, mostly focusing on the microphysics of the crustal lattice \citep{horowitz2008molecular, horowitz2009breaking, horowitz2011neutron, caplan2017colloquium, baiko2018breaking}. In particular \citet{horowitz2009breaking} found an exceptionally strong crust with breaking strain $\sigma \approx 0.1$.  They also found that the crust fails abruptly and universally, without cracking locally, because the pressure suppresses the formation of local imperfections, such as voids and dislocations. Such imperfections typically form in terrestrial materials and cause localised failure. The above breaking strain is much larger than previous estimates, e.g. $\sigma \sim 10 ^{-5}$ \citep{smoluchowski1970frequency}, which assume lower densities and do not use a long-range Yukawa-type interaction.

The results of \citet{horowitz2009breaking} are for $\sim 10^{11}$ cubic femtometres of material. In a real star it is likely that the macroscopic crust undergoes a different mode of failure than the microscopic material because of the presence of lattice dislocations, grain boundaries, vestiges of previous failures, and other mesoscopic im-perfections \citep{kittel1996introduction}. Even though the pressure is high, it is unlikely realistically to iron out these non-equilibrium features completely, and their existence seeds the production of additional imperfections, as the star spins down.

\subsection{Macroscopic redistribution and dissipation of stress due to failure}
\label{sec:Macro}
In this section, we present an idealised model for how the strain angle accumulated during spin down as in Section \ref{sec:SecDefNoFail} is dissipated and redistributed macroscopically following a failure event triggered microscopically as in Section \ref{sec:Micro}.

\begin{figure}
\includegraphics[height=5.5cm,width=8.5cm]{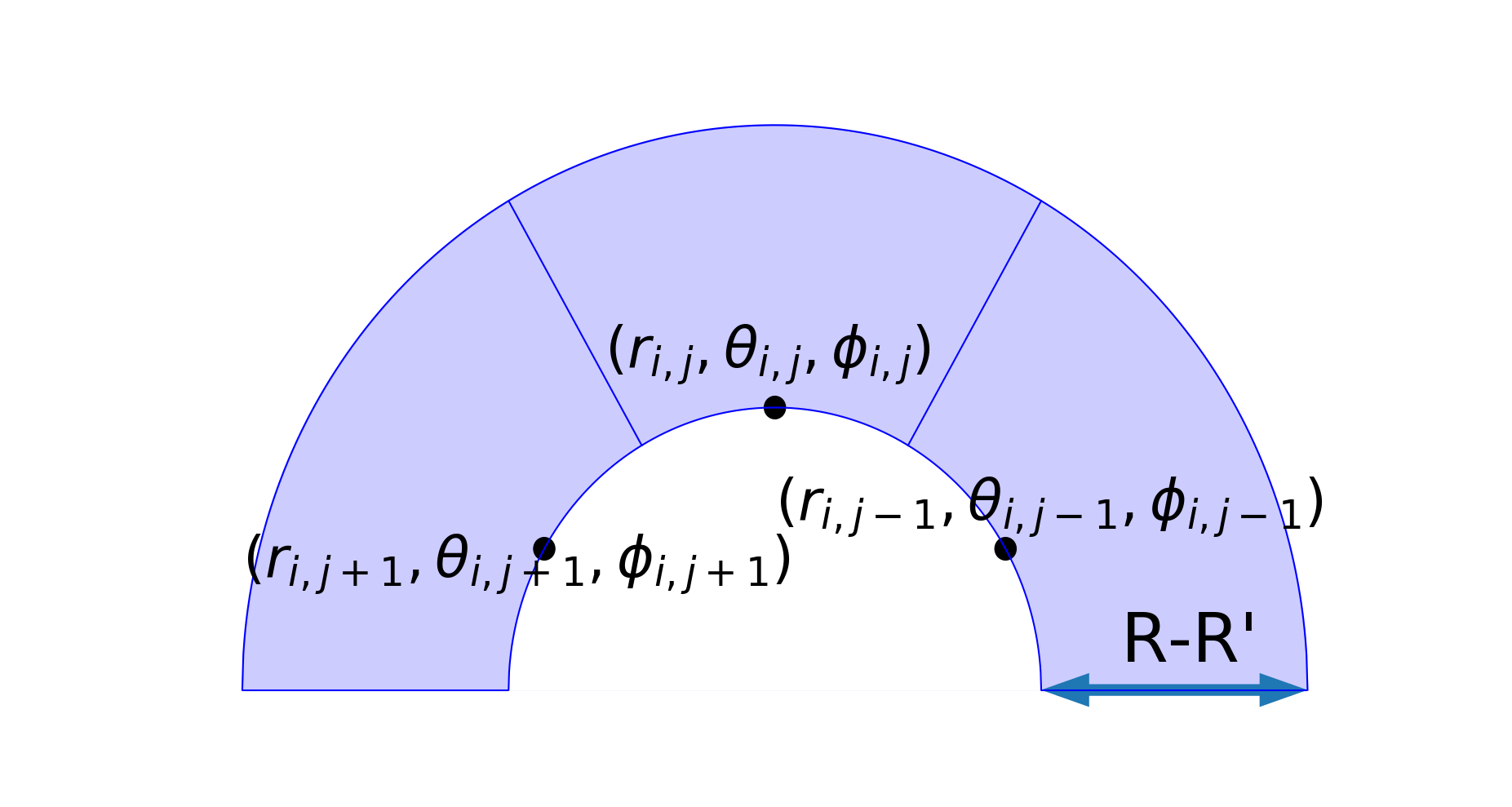}
\includegraphics[height=5.5cm,width=8.5cm]{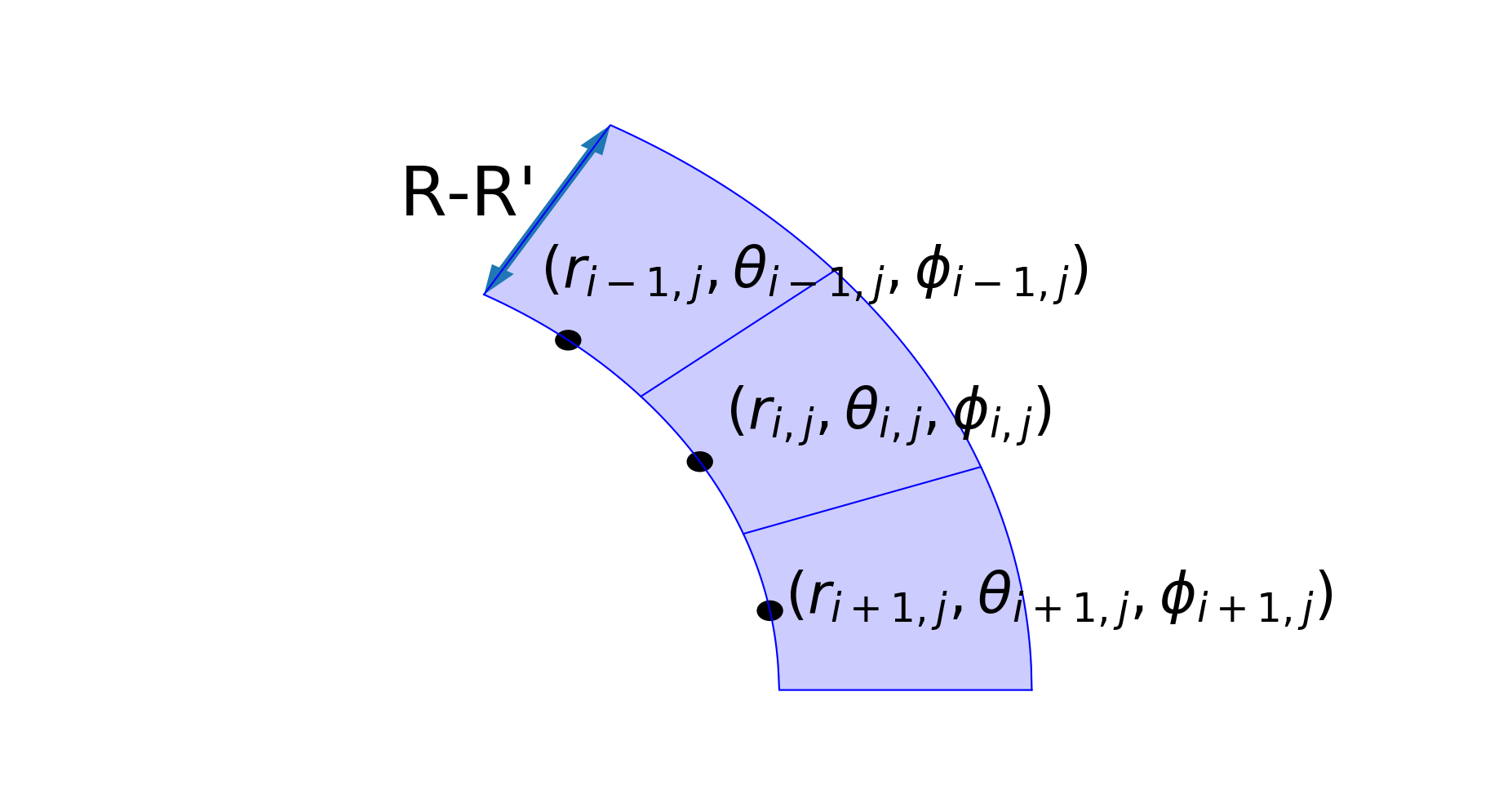}
\caption{Schematic diagram of cell geometry. (Top panel.) Top-down view, looking down the $z$-axis, showing the azimuthal variation with $\theta$ held constant. (Bottom panel.) Meridional, constant-$\phi$, cross-section of three adjacent cells. The black dots are the grid point coordinates assigned to a cell, which are used to calculate the strain increments and deformation vector $\vec{u}(\vec{r})$ during spin down. See Appendix A for additional context.}
\label{fig:CellVolumeDiagram}
\end{figure}

Failure on macroscopic scales is a discrete process in general. In this model we break up the crust into a two-dimensional grid of cells and track the strain angle in each cell. When the strain angle exceeds a critical threshold (i.e. when the cell fails) we dissipate and redistribute energy stored in the cell as mechanical potential energy. A given cell labelled with indices $i$ and $j$ ($0 \leq i,j \leq N-1)$ with coordinates $(r_{i,j},\theta_{i,j},\phi_{i,j} )$ occupies the volume defined by $r_{i,j} \leq r \leq r_{i,j}+(R-R')$, $\arccos\{[\cos(\theta_{i-1,j})+\cos(\theta_{i,j})]/2\}< \theta <\arccos\{[\cos(\theta_{i,j})+\cos(\theta_{i+1,j})]/2\}$, $(\phi_{i,j-1}+\phi_{i,j})/2 < \phi < (\phi_{i,j}+\phi_{i,j+1})/2$. This defines a lattice of $N$ ``rings'' of cells evenly spaced along the $z$-axis with each ``ring'' containing $N$ cells evenly spaced in $\phi$. In Fig. \ref{fig:CellVolumeDiagram} we display top-down and side-on diagrams of the cell volumes, in the upper and lower panels respectively. The black dots are the coordinates assigned to each cell. One can see that the cells are similar in shape to trapezoidal prisms with radial thickness $R-R'$. They are defined to encase the fluid core with no overlap at all times. More details, including an explanation of how overlap is prevented, are given in Appendix A. The deformation vector $\vec{u}_{i,j}$ of a cell and associated strain angle $\xi_{i,j}$ are determined as per Section \ref{sec:SecDefNoFail} and evaluated at $(r_{i,j},\theta_{i,j},\phi_{i,j})$. We evaluate $\vec{u}_{i,j}$ and $\xi_{i,j}$ at the crust-core boundary, because that is where the crust is densest and stiffest and where most of the elastic energy is located.

Each cell is assigned a random breaking strain in the range $0.075\leq i,j \leq 0.11$ informed by the polycrystalline stress-strain curve calculated by \citet{horowitz2009breaking}. Following \citet{fattoyev2018crust} we define failure to occur when $\gamma_{i,j} \geq \sigma_{i,j}$ is satisfied for some $(i,j)$, i.e. failure occurs when the strain angle in some cell exceeds its breaking strain. Other valid definitions of failure include the Tresca criterion, $\gamma_{i,j} \geq \sigma_{i,j}/2$ for some $(i,j)$, as used by \citet{giliberti2021starquakes} in the context of accreting neutron stars. We use the terms strain and strain angle interchangeably for simplicity's sake.

\begin{table}
\caption{Key energy components before, during, and after failure. The first two rows give the elastic energy in a failed cell just before and after failure respectively. The next seven rows quote the gravitational potential energy changes and heat dissipated while raising and lowering the failed cell and its neighbours. The last row gives the increase in elastic potential energy of a cell adjacent to one or more failed cells. In the right-hand column the fifth and sixth rows are equal as discussed in Sections \ref{sec:Macro} and \ref{sec:RadialMovement}. The parameters $0<A<1$ and $0<D<1$ are constants of the automaton.}
\begin{tabular}{c|c}
\hline
Energy Component & Formula\\
\hline
Elastic energy in failed  & $U_{i,j}^{\rm pre}=\frac{\mu\gamma_{i,j}^2V_{i,j}}{2}$\\
cell before failure & \\
 & \\
 
Elastic energy in failed  & $U_{i,j}^{\rm post}=AU_{i,j}^{\rm pre}$\\
cell after failure & \\
 & \\

Energy distributed from failed  & $E_{i,j}^{\rm dist}=D(1-A)U_{i,j}^{\rm pre}$\\
cell to neighbours & \\
 & \\

Plastic work to raise failed cell  & $\Delta E_{i,j}=(1-D)(1-A)U_{i,j}^{\rm pre}$\\
 & \\ 
 
Increase in gravitational potential & $(1-\beta)\Delta E_{i,j}$\\
energy of failed cell & \\
 & \\

Decrease in gravitational potential & $(1-\beta)\Delta E_{i,j}$\\
energy of lowered neighbours & \\
 & \\

Heat dissipated during plastic work & $\beta\Delta E_{i,j}$\\
to raise failed cells & \\
 & \\

Heat dissipated after & $(1-\beta)\Delta E_{i,j}$\\
lowering of neighbours  & \\
 & \\

Total heat dissipated & $\Delta E_{i,j}$\\
 & \\
 
Total energy received by & $E_{i,j}^{\rm NN}=\frac{1}{4}\sum_{i'=i\pm1 }E_{i',j}^{\rm dist}$\\
cell $(i,j)$ from failed neighbours & $\qquad\quad +\frac{1}{4}\sum_{j'=j\pm1 }E_{i,j'}^{\rm dist}$\\
\hline
\label{tab:EnergyComponents}
\end{tabular}

\end{table}

The elastic potential energy $U$ in a material of volume $V$ under strain $\gamma$ is given by the strain-energy formula
\begin{eqnarray}
\frac{\mu \gamma^2}{2}=\frac{U}{V}.
\label{eq:StrainEnergyFormula}
\end{eqnarray}
When a volume of material fails it partially relaxes. A portion of $U$ stays within the failed volume, while the rest does reversible mechanical work on neighbouring volumes, or does plastic work on the failed volume. Table \ref{tab:EnergyComponents} summarizes for the reader's convenience the key energy components which are stored, redistributed, and dissipated in and around a failed cell before, during, and after failure. We refer repeatedly to the formulas and notation in Table \ref{tab:EnergyComponents} throughout the rest of the paper.

A failed cell $(i,j)$ retains some fraction, $0<A<1$, of its energy after failure. This means that it retains some fraction of its strain as per
\begin{eqnarray}
\gamma_{i,j} \mapsto \sqrt{A}\gamma_{i,j}.
\label{eq:StrainUpdate}
\end{eqnarray}
In this paper $A$ is constant and takes the same value in every cell. When a volume of material fails the neighbouring material experiences an increase in strain in general, i.e. redistribution. The failed volume becomes less able to support mechanical loads, shifting the load onto the neighbouring material by doing reversible, mechanical work \citep{deformationandfracturemechanics}. Accordingly adjacent cells receive some reversible fraction, $E_{i,j}^{\rm dist}$, of the energy surrendered by the failed cell, $U_{i,j}^{\rm pre}-U_{i,j}^{\rm post}$, defined in the third row of Table \ref{tab:EnergyComponents} in terms of parameter $0<D<1$. In this paper $D$ is constant and takes the same value in every cell. Each adjacent cell $(i,j)$ receives an equal portion, $E_{i,j}^{\rm dist}$. \footnote{The cells closest to the poles $(i=0,N-1)$ have three neighbours rather than four. If one of these cells fails, energy is redistributed to cells $(0,j\pm1)$ and $(1, j)$ for $i=0$ or to cells $(N-1,j\pm1)$ and $(N-2,j)$ for $i=N-1$. Additionally cells with $j=0$ and $j=N-1$ are adjacent. This also modifies
the formula in the last row of Table \ref{tab:EnergyComponents}.} The strain of adjacent cells increases according to
\begin{eqnarray}
\gamma_{i\pm1,j} &\mapsto \sqrt{\gamma_{i\pm1,j}^2+\frac{2E^{\rm NN}_{i\pm1,j}}{\mu V_{i\pm1,j}}}, \label{eq:Diffusion1}\\
\gamma_{i,j\pm1} &\mapsto \sqrt{\gamma_{i,j\pm1}^2+\frac{2E^{\rm NN}_{i,j\pm1}}{\mu V_{i,j\pm1}}}, \label{eq:Diffusion2}
\end{eqnarray}
where $E^{\rm NN}_{i,j}$ is the \textit{total} energy distributed to cell $(i,j)$ by all of its failed neighbours.

The elastic energy decrement left over after redistribution to adjacent cells, namely $\Delta E_{i,j}=U_{i,j}^{\rm pre}-U_{i,j}^{\rm post}-E_{i,j}^{\rm dist}$ , goes into deforming the crust outwards against the force of gravity. The deformation is plastic and irreversible. In order to preserve hydrostatic equilibrium, the crust surrounding the failed cell must simultaneously move radially inwards such that the gravitational potential energy of the whole crust does not change; that is, the average radius of the entire crust does not change. The potential energy lost, when the crust moves inward through the centrifugal-gravitational potential, is dissipated as heat. A recipe for moving the crust inward and outward and calculating the plastic work done is presented in Section \ref{sec:RadialMovement}.

\subsection{Radial movement}
\label{sec:RadialMovement}

In response to the post-failure relaxation process described in Eqs. (\ref{eq:StrainUpdate}), (\ref{eq:Diffusion1}), and (\ref{eq:Diffusion2}), the failed cell $(i,j)$ moves outwards radially according to,
\begin{eqnarray}
r_{i,j} &\mapsto & r_{i,j}+\Delta r_{i,j},\label{eq:CellMovement}
\end{eqnarray}
with $\Delta r_{i,j}$ defined below in Eq. (\ref{eq:FailEnergy}). Its nearest neighbours move inward radially according to
\begin{eqnarray}
r_{i',j'} &\mapsto & r_{i',j'}+w_{i',j'},\label{eq:NeighbourMovement}
\end{eqnarray}
with $(i',j')=(i,j\pm1)$ and $(i',j')=(i\pm1,j)$ and $w_{i',j'}$ defined below in Eq. (\ref{eq:NeighbourFailEnergy}). The angular coordinates of each cell are held constant. $\Delta r_{i,j}$ is calculated by determining the elastic potential energy difference between the pre- and post-failure states of the crust (i.e. the first minus the second and third rows in Table \ref{tab:EnergyComponents}) and moving the failed cell $(i,j)$ the corresponding radial distance through the gravitational-centrifugal potential of the star. The total potential is given by \citep{franco2000quaking}
\begin{align}
\Phi(\vec{r}) &= -\pi G \rho_{\rm core} \left[2R^2 -\frac{2r^2}{3}-\frac{4e^2 r^2}{15}P_{2}(\cos\theta)\right],
\label{eq:Potential}
\end{align}
where $G$ is the gravitational constant, $\rho_{\rm core}$ is the density of the fluid core, and $e$ is the eccentricity of the ellipse defined by the meridional cross-section of the star. We calculate $\Delta r_{i,j} \ll r_{i,j}$ from
\begin{align}
\Delta r_{i,j}\rho_{\rm crust} V_{i,j} \frac{d\Phi(\vec{r})}{dr}\Big\vert_{\vec{r}=\vec{r}_{i,j}}=(1-\beta)\Delta E_{i,j},
\label{eq:FailEnergy}
\end{align}
where $\mu= 2.4 \times 10^{29}\,{\rm J m^{-3}}$ is the shear modulus of crust at the crust-core boundary as calculated by \cite{lander2015magnetically}, who used the methodology of \cite{horowitz2008molecular}, and the factor $(1-\beta)$ describes thermal losses during plastic deformation as discussed in Section \ref{sec:ThermalLosses} (fiducial value $\beta=0.9$).

As explained above, in order to preserve hydrostatic equilibrium, the cells adjacent to the failed cell must move inward such that the gravitational-centrifugal potential energy of the whole crust remains constant. Assuming for simplicity that the potential energy change is equal for every adjacent cell,\footnote{As described in Footnote 1 the cells closest to the poles have three neighbours rather than four. If one of these cells fails their neighbours move inwards per Eq. (\ref{eq:NeighbourFailEnergy}) but a factor of $1/3$ should be used rather than the factor of $1/4$ on the left-hand side of the equation.} we obtain
\begin{eqnarray}
\frac{1}{4}(\beta-1)\Delta E_{i,j} = w_{i',j'}\rho_{\rm crust}V_{i',j'}\frac{d\Phi(\vec{r})}{dr}\Big|_{\vec{r}=\vec{r}_{i',j'}},
\label{eq:NeighbourFailEnergy}
\end{eqnarray}
for the four cells$(i',j')=(i,j\pm1)$ and $(i',j')=(i\pm1,j)$. In other words the crust bends such that the centrifugal-gravitational potential energy of the whole crust is constant. The radius of the crust does not rise or fall on average. 

\begin{figure}
\includegraphics[height=5.5cm,width=8.5cm]{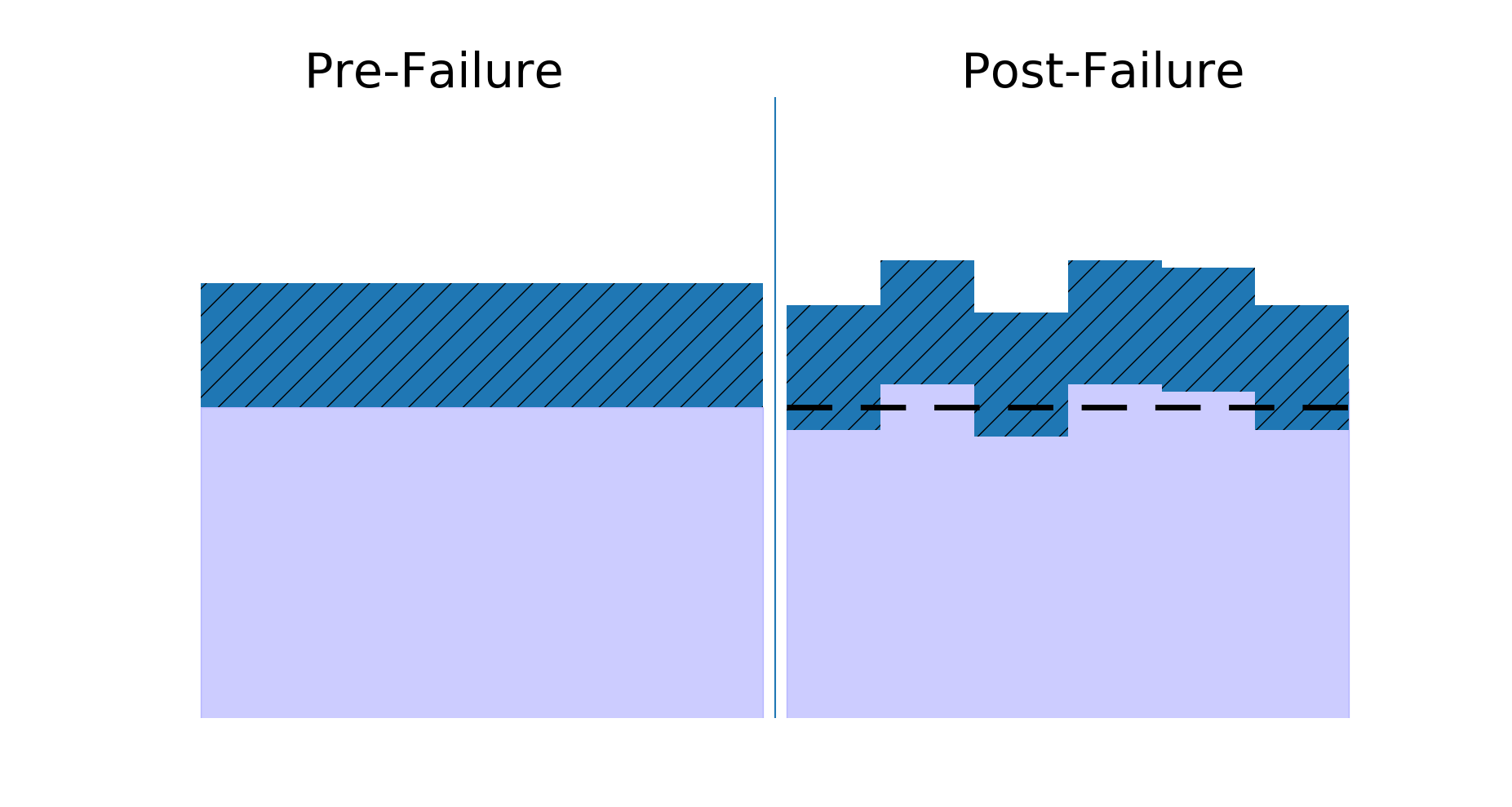}
\caption{Schematic view of a meridional cross-section of a segment of the crust and core pre- and post-failure. Cells move radially away from the core as they fail and the failed cell's neighbours move towards the core, such that the total change in gravitational potential energy of the raised and lowered cells is zero, as discussed in Sections \ref{sec:Macro} and \ref{sec:RadialMovement}. The amount of movement is greater for a more energetic failure, as seen in the uneven height in the second and third cells from the right-hand edge of the figure. Knock-on avalanches can occur when energy is redistributed to neighbouring cells. The black dashed line is the crust-core boundary of the pre-failure configuration. The diagram neglects the curvature of the star for clarity.}
\label{fig:FailureProfile}
\end{figure}

The statement that the failed and adjacent cells move outward and inward respectively is consistent with Eq. (\ref{eq:Potential}), because the gravitational-centrifugal potential increases strictly with $r$. Fig. \ref{fig:FailureProfile} illustrates how neighbouring cells lift or fall by different heights and produce an uneven solid surface. The crust is supported after lifting by a combination of fluid pressure from below and shear forces from adjacent regions of the crust.

When a cell fails its breaking strain resets in the above range, $0.075 \leq \sigma_{i,j} \leq 0.11$. The microscopic change of the crystal structure after experiencing failure is still a matter of investigation \citep{baiko2018breaking, chugunov2020neutron, kozhberov2020breaking} and we do not investigate it here. We assume that the crystal structure relaxes to a similar pre-failure configuration (albeit under a different strain) on a timescale much shorter than the waiting time until the next failure.

\subsection{Thermal losses during plastic deformation}
\label{sec:ThermalLosses}
When a body undergoes plastic deformation a fraction, $\beta$, of the plastic work done on the body is converted into heat. The right-hand side of Eq. (\ref{eq:FailEnergy}) gives the reversible energy transfer i.e. the energy not lost as heat during plastic deformation. Replacing the factor $1-\beta$ with $\beta$ gives the energy that is lost as heat. Together with the heat released by the neighbouring cells moving inward, the total heat dissipated by crustal movement (triggered by failed cells) in a particular time-step is denoted by $\Delta E$ in what follows. It is obtained by summing row four (equivalent to row nine) in Table \ref{tab:EnergyComponents} over all cells that fail in the time-step.

The exact value of $\beta$ depends on the specific material in question and a variety of factors such as strain, strain rate and deformation history \citep{rittel1999conversion, rosakis2000thermodynamic, macdougall2000determination, hodowany2000partition, ravichandran2002conversion}. Appendix B outlines briefly some of the key issues. It is challenging to predict $\beta$ from first principles even for terrestrial materials, where one is guided by controlled experiments, let alone for neutron star matter. Common terrestrial metals like copper and steel are measured to have $0.75\lesssim \beta \lesssim 0.95$, reflecting a competition between internal friction and the formation of lattice defects. In the face of this uncertainty, we adopt $\beta=0.9$ in this paper, taking the conservative position astrophysically that most of the elastic energy released during plastic deformation is dissipated as heat and does not contribute to mountain formation. Hence the mass ellipticity and gravitational wave signal (which are proportional roughly to $\beta$) are smaller than one would predict otherwise. The reader is encouraged to experiment with other values. We reiterate that we neglect magnetic forces in this paper, in order to focus on the new idea of tectonic activity modelled by a cellular automaton. Hence the energy lost during plastic deformation is due to internal friction and vibrations, not due to rearranging the magnetic field \citep{srinivasan1990novel, ruderman1991neutron1, ruderman1991neutron2, ruderman1991neutron3}.

\subsection{Cellular automaton}
\label{sec:CellAutomaton}
We now combine the idealised tectonic physics of deformation and failure in Sections \ref{sec:SecDefNoFail}, \ref{sec:Macro}, \ref{sec:RadialMovement} and \ref{sec:ThermalLosses} into a simple automaton, which is iterated to study the long-term, far-from-equilibrium dynamics of the system. The state of each cell of the automaton is described by five numbers, $[r_{i,j}(t_{n}),\theta_{i,j}(t_{n}),\phi_{i,j}(t_{n}),\gamma_{i,j}(t_{n}),\sigma_{i,j}(t_{n})]$, which are updated at each time-step. Here the $i$ and $j$ indices $(0\leq i,j \leq N-1)$ refer to the cell index and $t_{n}$ corresponds to the time-step.

The automaton is initialised as a Maclaurin spheroid with eccentricity $e=0.1$, a representative value. The oblateness at birth is uncertain, with estimates reaching as high as $e\approx0.6$. \citep{haensel2002equation}. For $e\ll1$ at birth and before failure we have \citep{franco2000quaking, haskell2008modelling}

\begin{eqnarray}
r_{i,j}(\theta_{i,j})&=&R' \left[1-\frac{e^2P_{2}(\cos\theta_{i,j})}{3} \right],\label{eq:Initialise1}\\
\theta_{i,j}&=&\arccos\Big(1-\frac{1}{N}-\frac{2i}{N} \Big)\label{eq:Initialise2}\\
\phi_{i,j}&=&\frac{2\pi j}{N}
\label{eq:Initialise3}
\end{eqnarray}

The cellular automaton entails the following steps:
\begin{enumerate}[label=\arabic*)]
		  \item Choose $A$ and $D$.
          \item Initialise $r_{i,j}(t_0)$, $\theta_{i,j}(t_0)$ and $\phi_{i,j}(t_{0})$ according to Eqs. (\ref{eq:Initialise1}), (\ref{eq:Initialise2}),  and (\ref{eq:Initialise3}).
          \item Assign to each cell a random breaking strain $0.075 \leq \sigma_{i,j} \leq 0.11$ and $\gamma_{i,j}=0$. 
          \item Calculate $\vec{u}(\vec{r}_{i,j})$ from Eqs. (\ref{eq:defvecs1})--(\ref{eq:Tensor}) for each cell for a small constant angular velocity decrement $\delta \Omega$ and update $\vec{r}_{i,j}(t_n)=\vec{r}_{i,j}(t_{n-1})+\vec{u}[\vec{r}_{i,j}(t_{n-1})]$. See Appendix A for further details on the effect of spin-down deformation on cell boundaries.
          \item Calculate $\alpha_{lm}(\vec{u}_{i,j})$ and $\xi(\vec{r}_{i,j})$ for each cell. Set $\gamma_{i,j}(t_{n})=\gamma_{i,j}(t_{n-1})+\xi(\vec{r}_{i,j})$.
          \item If cell $(i,j)$ has $\gamma_{i,j} \geq \sigma_{i,j}$ it undergoes failure. 
          \begin{enumerate}[label=\alph*)]
          	\item Increase the strain adjacent to cell $(i,j)$ according to  Eqs. (\ref{eq:Diffusion1}) and (\ref{eq:Diffusion2}) synchronously, i.e. all cells are tested for failure first, then cells with $\gamma_{i,j}\geq\sigma_{i,j}$ fail together, before the strains of neighbouring cells are increased.
          \item Update the coordinates of cell $(i,j)$ and its neighbours according to Eqs. (\ref{eq:CellMovement})--(\ref{eq:NeighbourFailEnergy}).
          \item Increase $\Delta E$, as described in the first paragraph of Section \ref{sec:ThermalLosses}.
          \item Reset $\gamma_{i,j}$ in the cell $i,j$ according to Eq. (\ref{eq:StrainUpdate}).
          \item Reset the breaking strain randomly within the range $0.075 \leq \sigma_{i,j} \leq 0.11$.
          \item Repeat the steps (6)(a) to (6)(f) until all cells satisfy $\gamma_{i,j} < \sigma_{i,j}$. 
          \end{enumerate}
          \item Decrease $\Omega$ by $\delta\Omega$, where $\delta\Omega$ is constant throughout the simulation.
          \item Repeat from step (4) until $\Omega$ drops to $\Omega=\delta\Omega$.
\end{enumerate}

We assume that the star spins down in response to a magnetic dipole torque satisfying $\dot{\Omega} \propto -\Omega^3$, and hence
\begin{align}
\Omega(t)&=\Omega(0) \left( 1+\frac{t}{\tau} \right)^{-1/2},
\label{eq:time}
\end{align}
with 
\begin{align}
\tau&=-\frac{\Omega(0)}{2\dot{\Omega}(0)}.
 \end{align}
Given $\Omega(0)$ and $\dot{\Omega}(0)$ we can calculate the age of the star from its current angular velocity. In this paper we assume that the time-scale of failure events is much shorter than the spin-down time-scale; failure is effectively instantaneous.

In this paper we simulate a star with the following fiducial parameters: $R=10.5\,$km, $R'=9.5\,$km, $\rho_{\rm crust}=10^{17}$ kgm\textsuperscript{-3} \citep{haensel2002equation, chamel2008physics}, $\rho_{\rm core}=6.38\times10^{17}$ kgm\textsuperscript{-3}, $\mu=2.4 \times 10^{29}\,{\rm J m^{-3}}$, $\Omega(0)/2\pi=800$ Hz, $\dot{\Omega}(0)/2\pi=1\times 10^{-8}$ Hzs\textsuperscript{-1}, and therefore $\tau=4\times10^{10}$ s. For these values the ratio $\Omega/\dot{\Omega}$ is in line with typical pulsars. Again, the values above are illustrative only; their uncertainties are outweighed by uncertainties in other material properties, such as $\beta$, $A$, and $D$. For $(A,D)=(0.5,0.5)$ and the above fiducial values Eq. (\ref{eq:FailEnergy}) implies $\Delta r_{i,j}\approx 0.16$mm for a failure event at $\sigma_{i,j}=0.1$ on the equator.  

\section{Failure event statistics}
\label{sec:Failureeventstatisitcs}
Starquakes triggered by failure events have been suggested as important factors	driving the phenomena of glitches \citep{ruderman1976crust, franco2000quaking, middleditch2006predicting, chugunov2010breaking, fattoyev2018crust,giliberti2020modelling, bransgrove2020quake} and gamma-ray bursts \citep{kaplan2001hubble, hurley2005exceptionally, horowitz2008molecular}. Recent theoretical work also shows that the long-term statistics of avalanche processes in neutron stars, including starquakes, can be understood in terms of well-defined stochastic mechanisms describing stress accumulation and release, e.g. state-dependent Poisson processes \citep{melatos2008avalanche, warszawski2013knock, fulgenzi2017radio, carlin2019generating, carlin2019autocorrelations} and Brownian threshold processes \citep{carlin2020long}. Models of these mechanisms make specific, falsifiable predictions about the long-term probability distribution functions (PDFs) and cross- and autocorrrelations of event sizes and waiting times and are therefore a promising tool for distinguishing between classes of microphysical processes, including crustal failure in the starquake context \citep{fulgenzi2017radio, melatos2018size, carlin2019generating, carlin2019autocorrelations, carlin2020long}. We are therefore motivated to investigate the statistics of failure events in this paper, partly as input into the foregoing studies.

We define a failure event as occurring at a particular time-step, when one or more individual cells fail during that time-step. Our definition is therefore expressed collectively for simplicity: for the purposes of terminology, a time-step corresponds to either zero or one global failure events, no matter how many individual cells fail, and whether or not the failed cells are contiguous (i.e. single or multiple disjoint avalanches). The size of a failure event ($\Delta E$, measured in units of joules) is defined as described in the first paragraph of Section \ref{sec:ThermalLosses}. The waiting time ($\Delta t$, measured in units of $\tau$) is the time between one event and the next.

This section is structured as follows. In Section \ref{sec:SizeWaitDistrib} we present representative examples of the size and waiting-time distributions and comment on their general form. In Section \ref{sec:SizeWaitCorr} we calculate size and waiting time cross-correlations, which may offer interesting opportunities to falsify crust failure models observationally. Section \ref{sec:History} explores how the distribution of failure events depends on the redistribution and dissipation parameters $A$ and $D$. The values of $A$ and $D$ also control the long-term level of tectonic activity in the star, which is studied in Section \ref{sec:RedistribDiss}.

\subsection{Size and waiting-time distributions}
\label{sec:SizeWaitDistrib}
\begin{figure}
\includegraphics[height=5.5cm,width=8.5cm]{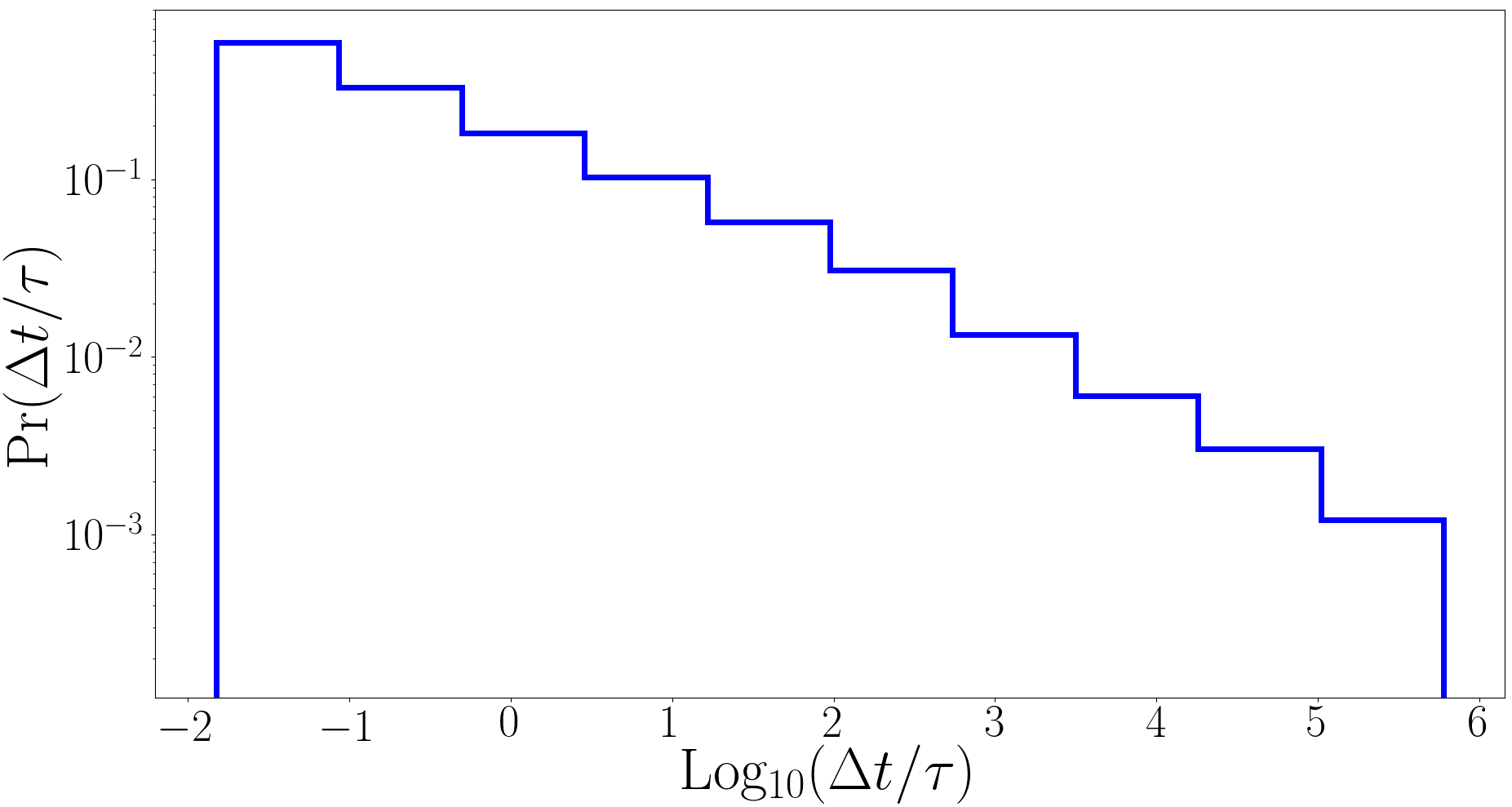}
\includegraphics[height=5.5cm,width=8.5cm]{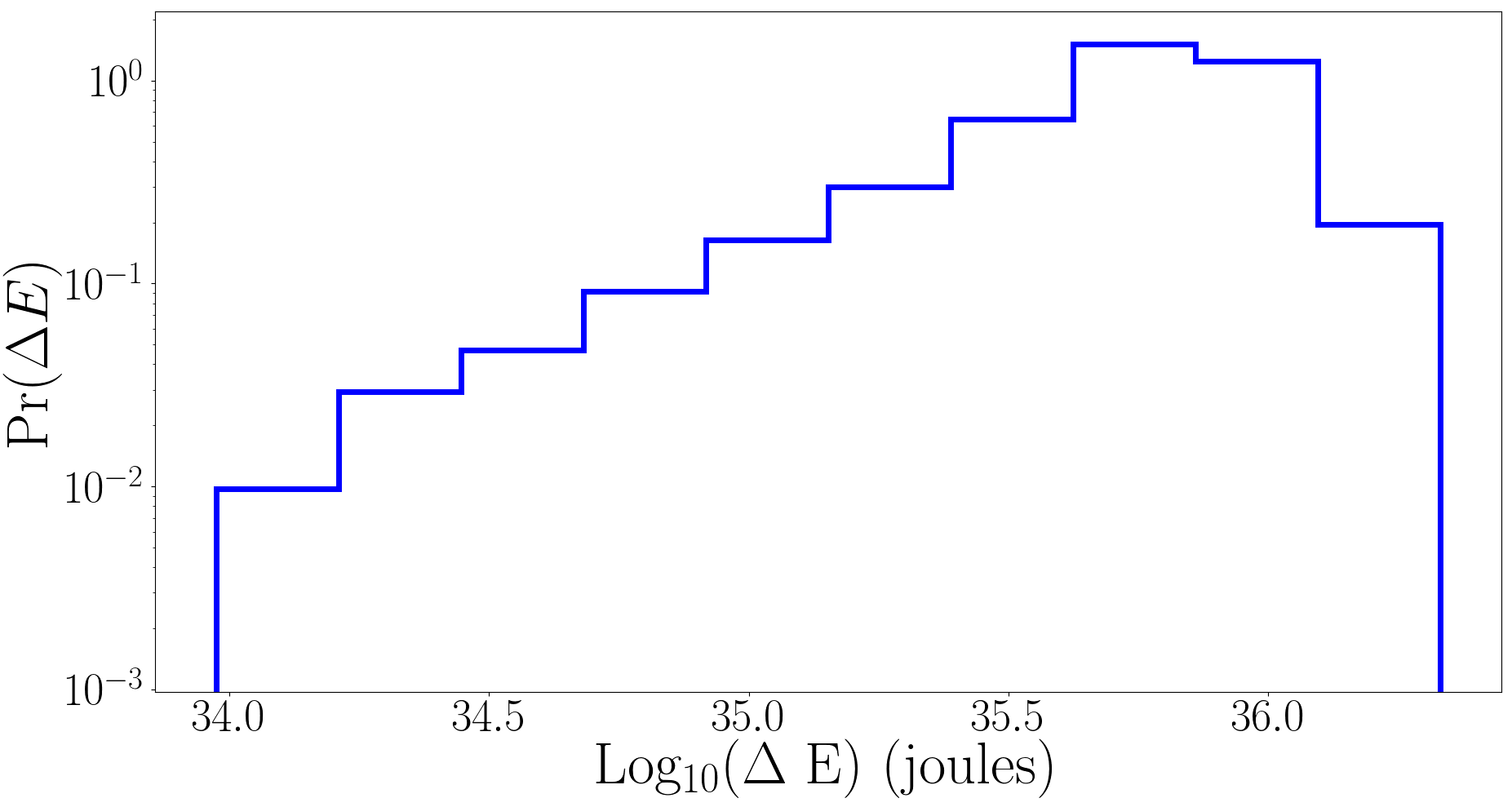}
\caption{Long-term failure statistics for $(A,D)=(0.5,0.5)$ and the fiducial
values noted at the end of Section \ref{sec:CellAutomaton}. (Top panel.) Waiting-time PDF in units of $\tau$. (Bottom Panel.) Event size PDF in units of joules. These PDFs are constructed from the same data comprising all the events from five simulations. The properties of these PDFs are examined in Section \ref{sec:SizeWaitDistrib} and how they change with respect to $A$ and $D$ is examined in Section \ref{sec:RedistribDiss}. These PDFs are constructed with $N=200$, and $\delta \Omega=1$.}
\label{fig:illustrative}
\end{figure}
We start by presenting the PDFs of the event sizes and waiting times for representative values of the failure parameters, $(A,D)=(0.5,0.5)$. They are displayed in Fig. \ref{fig:illustrative}. The PDFs are qualitatively similar across the relevant domain of $A$ and $D$ (see Section \ref{sec:RedistribDiss}). The waiting-time PDF (upper panel of Fig. \ref{fig:illustrative}) rises steeply at low $\Delta t$ and has an approximately power-law tail at high $\Delta t$. The peak occurs at $\Delta t \sim 10^{-2}$ in units of $\tau$. However a reasonable number of failures occur with shorter and longer ($\gtrsim 10\tau$) waiting times. The size PDF (lower panel of Fig. \ref{fig:illustrative}) is more tightly peaked, reaching a maximum at $\Delta E \approx 5.5 \times 10^{35}$ J with a full-width-at-half-maximum of approximately $10^{36}$ J.

We observe that the waiting-time PDF is approximately a power law with a cut-off at low $\Delta t$. The cut-off is an artefact of how time is discretised in the automaton: $\delta \Omega$ is fixed and therefore sets a minimum time increment and hence a minimum $\Delta t$. The power-law distribution is evidence of a non-Poissonian point process governing failure events, consistent with other self-organized critical systems \citep{jensen1998self}. The reader is referred to Appendix C for further details.

Regarding the size PDF, the size of an event depends, in part, on the volume of the failed cells. This affects the location of the cut-off at low $\Delta E$. In the lower panel of Fig. \ref{fig:illustrative} the cut-off in $\Delta E$ is $\approx 1\times10^{34} J$. For $N=200$ and a cell at the equator failing at $\sigma_{i,j} = 0.10$, we have $\Delta E \approx 8.4\times 10^{33}$J from Table \ref{tab:EnergyComponents}, in good agreement with the observed cut off. In contrast, the largest events correspond to avalanches involving large numbers of cells. Their sizes are unaffected by the total number and volume of cells in the automaton. With the peak in the size PDF occurring at $\approx 4\times 10^{35}$J, the most common events involve $\approx 50$ cells.

Different events have different sizes due to the random breaking strains, $\sigma_{i,j}$, and the random number of cells failing during one time-step, $N_{\rm fail}$. It is interesting to check how much each of these two factors affects the dispersion of event sizes, where ``dispersion'' here is a general term for the characteristic width of the size PDF, which does not necessarily equal exactly the full-width-half-maximum or standard deviation. The failure energy, $\Delta E$, depends quadratically on $\sigma_{i,j}$, which is randomly assigned to each cell as per Section \ref{sec:CellAutomaton}. Comparing the mean of the square and the square of the mean of the $\sigma_{i,j}^2$, $N_{\rm fail}$ and $\Delta E$ distributions, we obtain $\langle \sigma_{i,j}^{4} \rangle^{1/2} \approx 1.02\langle \sigma_{i,j}^2 \rangle$, $\langle N_{\rm fail}^2 \rangle ^{1/2}\approx 1.12\langle N_{\rm fail} \rangle$ and $\langle \Delta E^2 \rangle^{1/2} \approx 1.14\langle \Delta E\rangle$ for $(A,D)=(0.5,0.5)$. This implies that the $\Delta E$ distribution is, in relative terms, approximately as broad as the $N_{\rm fail}$ distribution and broader than the $\sigma_{i,j}^2$ distribution. Hence the stochasticity of avalanches (i.e. $N_{\rm fail}$) explains somewhat more of the dispersion in event sizes than the random breaking strains of individual cells.

We find that the number of events is linearly proportional to the number of cells in the automaton, and $\Delta E$ is inversely proportional to the number of cells, but the observable, macroscopic work lost as heat ($\Sigma \Delta E$) is unaffected by the grid size or the frequency decrement. The volume of the crust that undergoes failure also converges with $N$ and $\delta \Omega$. Appendix D investigates the convergence of
the automaton.

\subsection{Size-waiting-time correlations}
\label{sec:SizeWaitCorr}
\begin{figure}
\includegraphics[height=5.5cm,width=8.5cm]{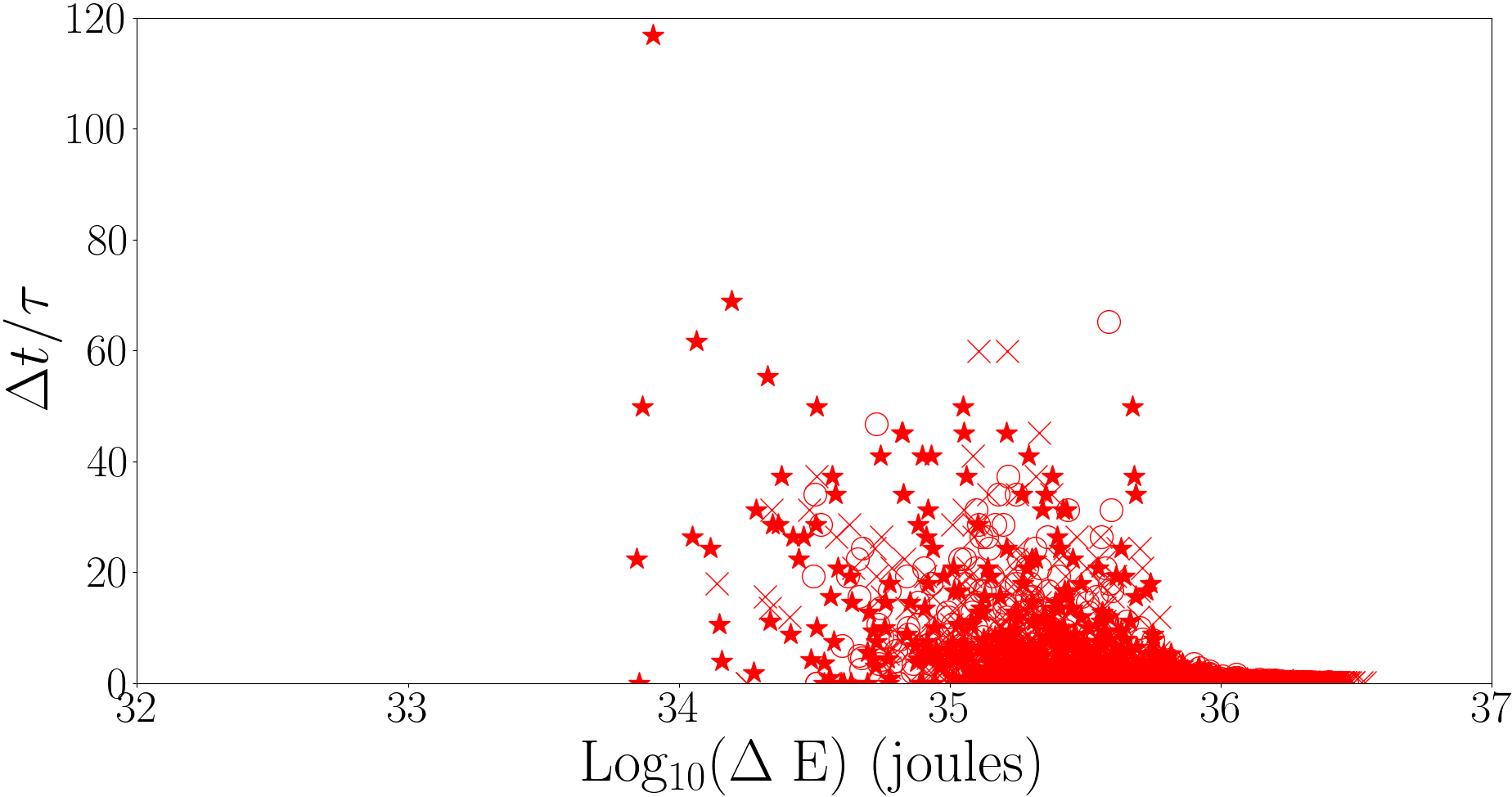}
\includegraphics[height=5.5cm,width=8.5cm]{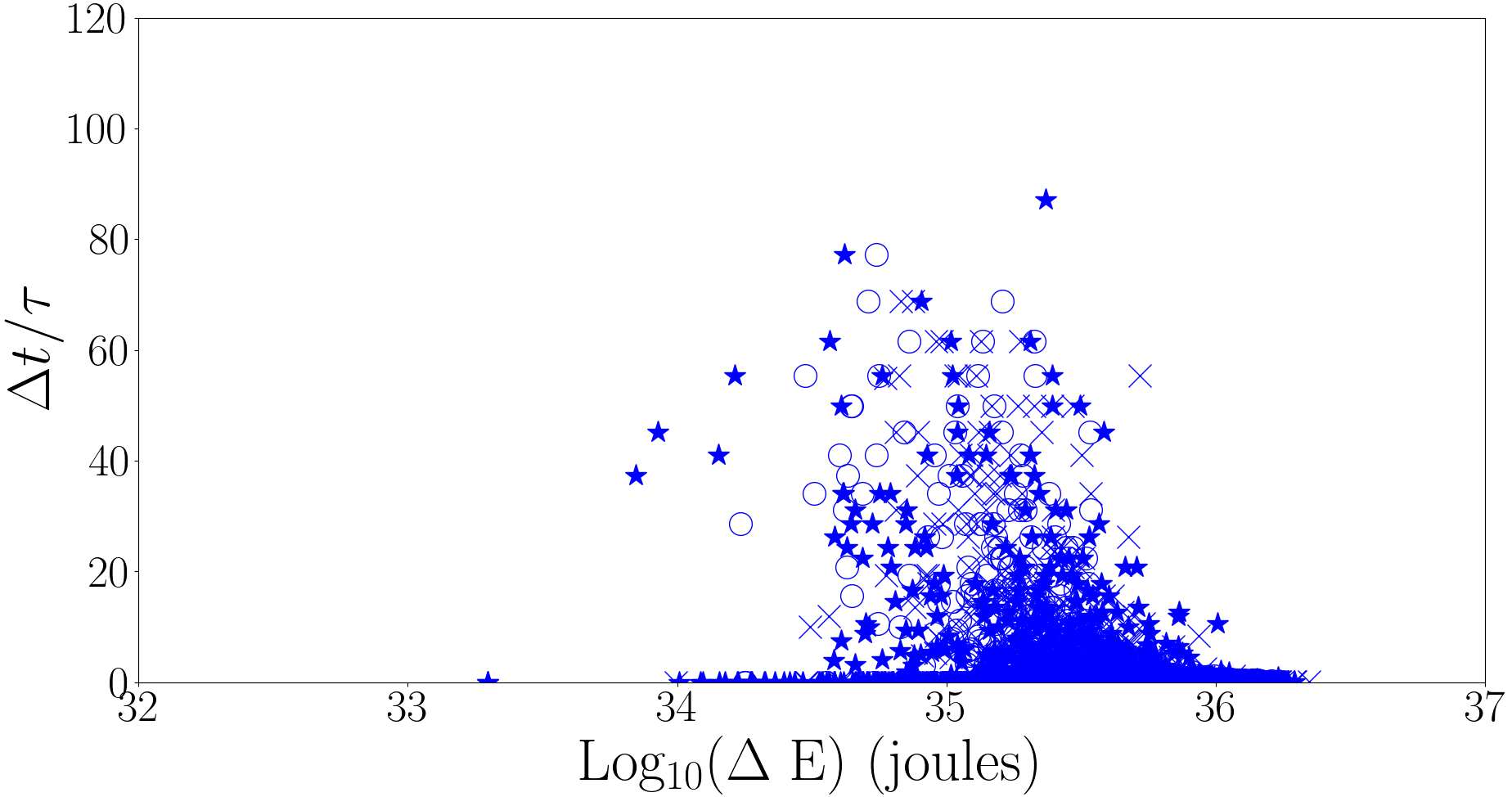}
\includegraphics[height=5.5cm,width=8.5cm]{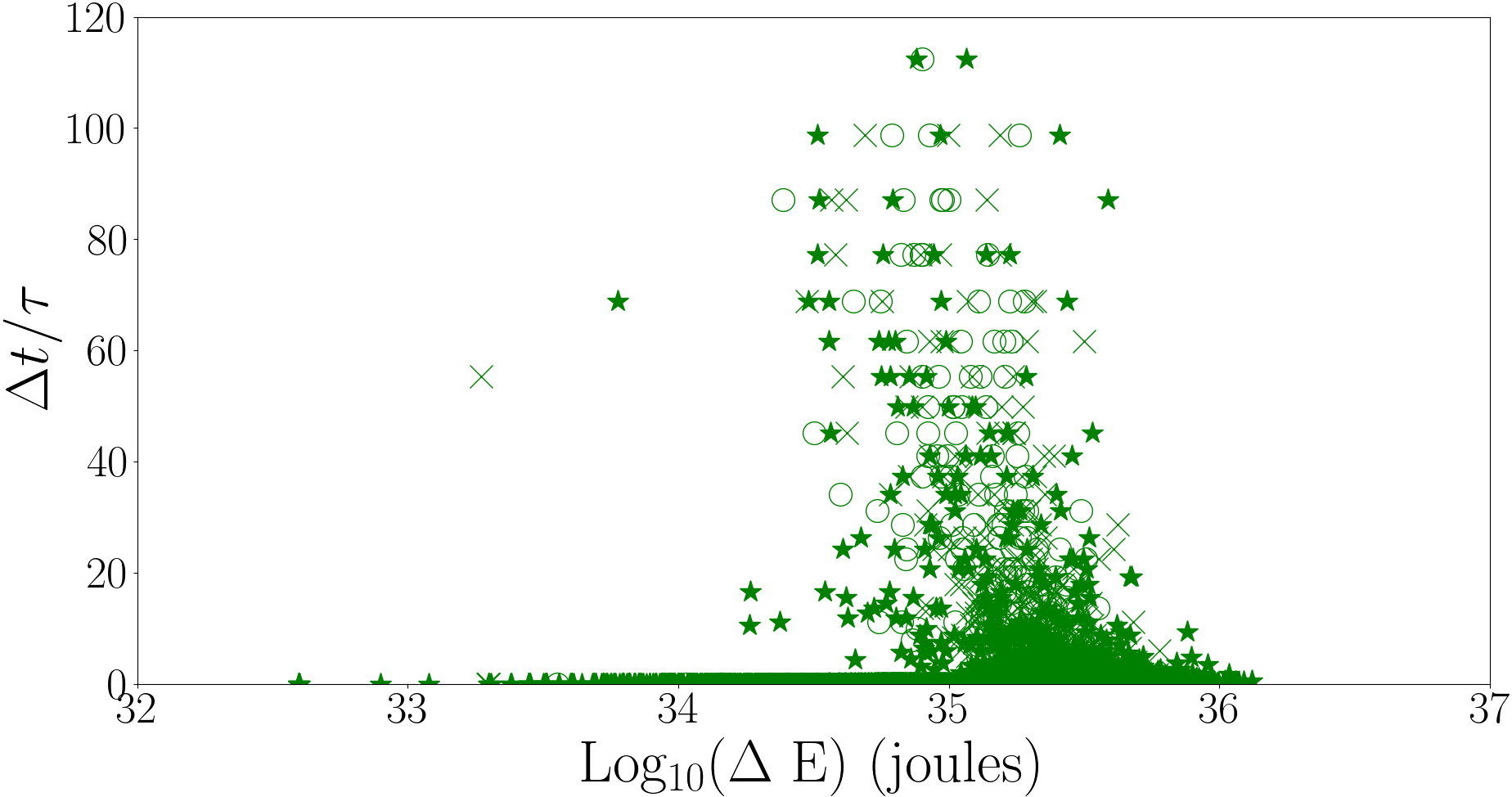}
\caption{Size-waiting-time correlations. Each data point marks the energy of a particular event and the time until the next event. Only the first $95\%$ of events in a star's life are included as the last events often have $\Delta t \gg 10^3$. Colours: red corresponds to $A = 0.1$ (upper panel), blue to $A = 0.5$ (middle panel) and green to $A = 0.9$ (lower panel). Shapes: empty circles correspond to $D = 0.1$, crosses to $D = 0.5$, and stars to $D = 0.9$. Data comes from five simulations per choice of $(A, D)$.}
\label{fig:SizeWaitDistrib}
\end{figure}
In Fig. \ref{fig:SizeWaitDistrib} we plot $\Delta E$ vs $\Delta t$ for individual events. Events cluster within ranges of $\Delta E$ with tails in $\Delta t$. The central value of $\Delta E$ in each cluster depends on $A$ and $D$. We find strong evidence for a correlation between $\Delta E$ and $\Delta t$ with the direction and strength of the correlation dependent on the value of $A$ but less sensitive to $D$. For $A = 0.9$ we find a Spearman rank coefficient of $\approx 0.3$ with a p-value $\leq 10^{-30}$. At $A = 0.5$ the Spearman rank coefficient is $\approx -0.2$ with p-value $\leq 10^{-2}$, and at $A = 0.1$ the Spearman rank coefficient is $\approx -0.6$ with p-value $\leq10^{-150}$. The Spearman rank coefficients are calculated with data from five realisations of the
automaton, with approximately 400 events per realisation per value
of $(A,D)$.

The positive correlation at large $A$ is due to the secular deformation of the crust. Spin-down deformation moves the crust away from the equator and towards the poles (see Fig. \ref{fig:PolarSpread} in Appendix A, alternatively see Fig. 2 in \cite{franco2000quaking}), causing the equatorial cells to grow as the star ages. Events later in the star's life are larger on average because the equatorial cells, where strain is greatest, have greater volume. Additionally strain accumulates slower, as the star ages, causing waiting times to be longer when the star is older. Later failures tend to be larger due to the larger equatorial cells, and have longer waiting times because the star is older, hence the positive correlation between event size and waiting time for large $A$. However for small $A$ later events are smaller than earlier ones: a cell failing at late times finds most of its neighbours have already failed, are strained below $\sigma_{i,j}$ and so are unlikely to be part of an avalanche. Later events being smaller than early events leads to a negative correlation between event size and waiting time. Equatorial cell volumes can increase by a factor of $\lesssim 2$ due to spin-down deformation, but a failed cell has up to four neighbours that may be part of an avalanche (and neighbours have neighbours). Hence the magnitude of the correlation is greater for small $A$ than for large $A$. Furthermore, by examining Fig. \ref{fig:EventsOverTime} in Section \ref{sec:History}, we see that heat is dissipated quickly for small $A$ but takes longer for large $A$. Large events are disfavoured at late times for small $A$ but not for large $A$.

Some events per run ($\approx 2\%$ of all events) have long waiting times $\Delta t/\tau \gtrsim 10^4$. These occur late in the star's life $(10^4 \lesssim t/\tau \lesssim 10^5$ , when the star is spinning at $\lesssim 10$Hz), when the strain build up has slowed considerably. These events are left out of Fig. \ref{fig:SizeWaitDistrib} for clarity but are included in the calculation of the Spearman rank coefficients.

\subsection{History of activity and energy release}
\label{sec:History}
\begin{figure}
\includegraphics[height=5cm,width=8.5cm]{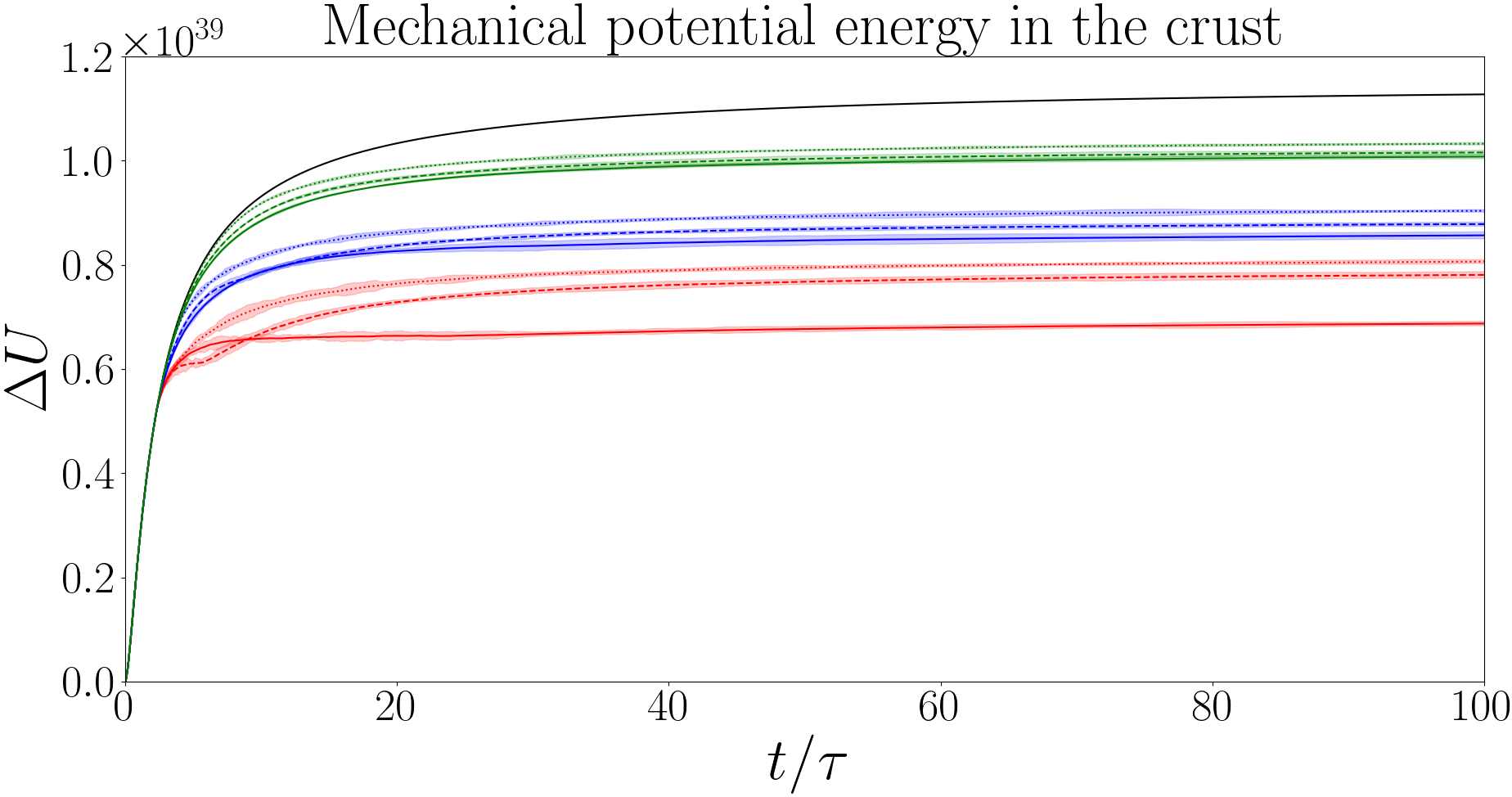}
\includegraphics[height=5cm,width=8.5cm]{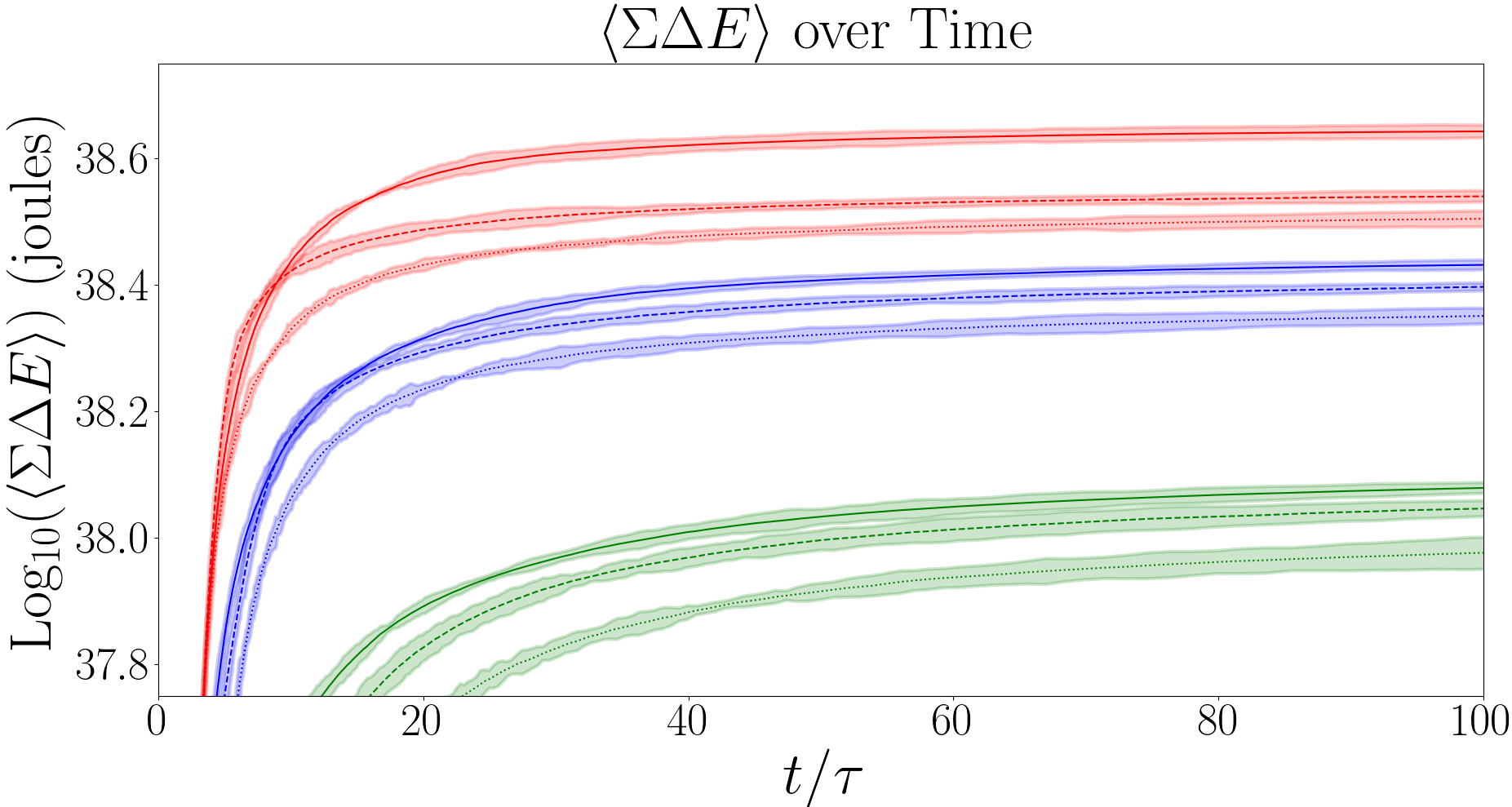}
\includegraphics[height=5cm,width=8.5cm]{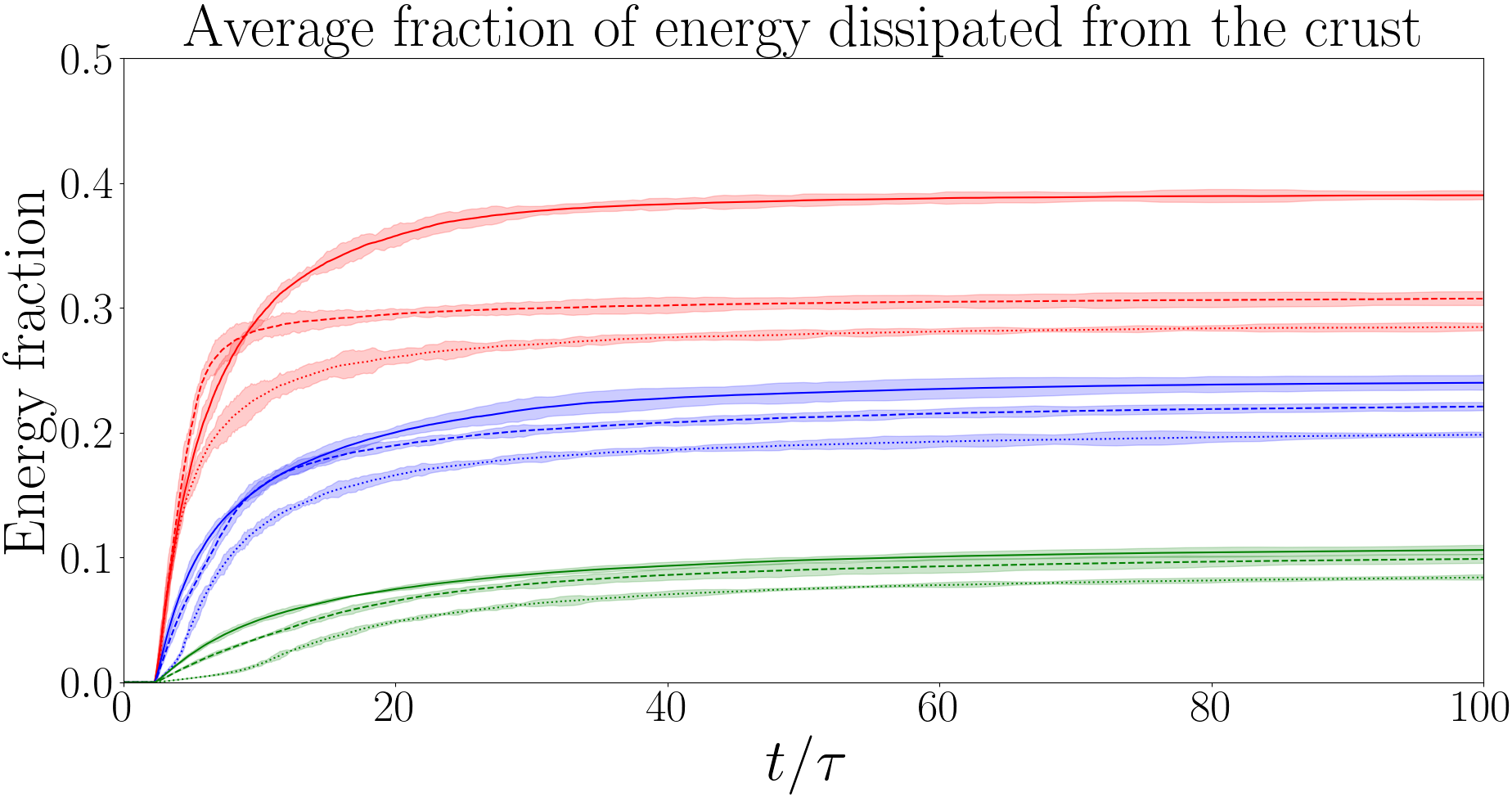}
\caption{Volume aggregated energetics over time. (Top panel.) The total mechanical potential energy stored in the crust at time $t/\tau$. The solid black line is the energy that would be stored in the crust if the crust never failed. (Middle panel.) Total heat dissipated (in units of joules) up to time $t/\tau$. (Bottom panel.) Heat dissipated during failure events up to time $t/\tau$ divided by the total energy deposited into the crust by spin down up to time $t/\tau$. All plots are the average results of five simulations per value of $(A, D)$. The shading corresponds to five standard deviations. Red corresponds to $A = 0.1$, blue to $A = 0.5$ and green to $A = 0.9$. The solid lines correspond to $D = 0.1$, dashed to $D = 0.5$, dotted to $D = 0.9$.}
\label{fig:DeltaUplots}
\end{figure}
\begin{figure}
\includegraphics[height=5cm,width=8.5cm]{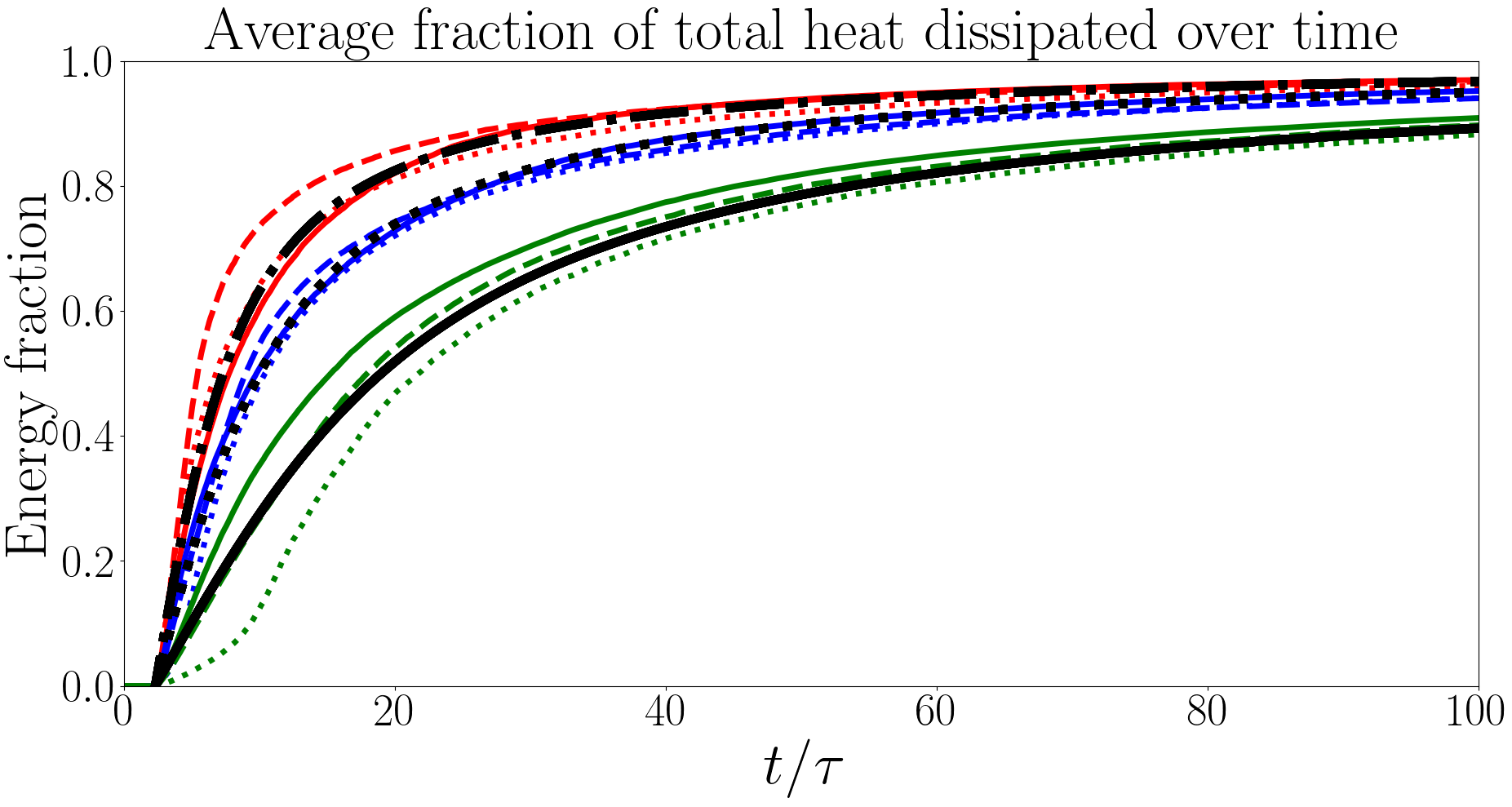}
\caption{
Failure energetics as a function of time. Heat dissipated during failure events up to time $t/\tau$ normalised by the total heat dissipated during all failure events over the star's life, i.e. the fraction of total heat dissipated up to time $t/\tau$. The black lines indicate an arbitrary, empirical fit $(2/\pi) \arctan[(0.22-0.18A)(t/\tau-2.3)]$, where the solid line corresponds to $A = 0.9$, dotted $A = 0.5$ and dot-dashed $A = 0.1$. All plots are the average results of five simulations per value of $(A, D)$. Red corresponds to $A = 0.1$, blue to $A = 0.5$ and green to $A = 0.9$. The solid lines correspond to $D = 0.1$, dashed to $D = 0.5$, dotted to $D = 0.9$.}
\label{fig:EventsOverTime}
\end{figure}

In previous sections we consider the size of events, where the size of an event is defined as the total heat dissipated by all failed cells, contiguous or not, in the time-step of the event. We are also interested in the total heat dissipated by all events up to an arbitrary instant in the star's life, i.e. the total energy released since birth, by failure, from the crust and dissipated as heat. The heat dissipated from crustal failure is relevant to a variety of astrophysical mechanisms that may connect to crustal failure such as glitches \citep{pines1972microquakes, ruderman1976crust} or fast radio bursts \citep{suvorov2019young}.

It has been suggested that some energetic transient phenomena such as gamma-ray bursts \citep{kaplan2001hubble, hurley2005exceptionally, horowitz2008molecular} originate in the breaking crust of neutron stars. In this and similar contexts, it is interesting to ask how much mechanical potential energy is stored in the crust and how much of the total energy deposited into the crust is released as heat. In the upper panel of Fig. \ref{fig:DeltaUplots} we display the total mechanical potential energy stored in the crust at time $t$ for a variety of values of $(A,D)$. In the middle panel we display the cumulative energy dissipated as heat ($\Sigma \Delta E$) in all failure events up to time $t$. In the lower panel we display the fraction of energy deposited into the crust that is dissipated as heat, i.e. the data of the middle panel divided by the black line of the upper panel. For the fiducial parameters listed at the end of Section \ref{sec:CellAutomaton} approximately $1.15 \times 10^{39}$J are deposited into the crust in the form of elastic potential energy over the star's life and between $\approx 1 \times 10^{38}$J and $\approx 5\times 10^{38}$J of heat are dissipated in all failure events. Between $\approx8\%$ and $\approx40\%$ of the deposited energy is eventually lost as heat depending on $A$ and $D$. The star releases the most heat near $(A,D)=(0,0)$ and the least near $(A,D)=(1.1)$. The qualitative behaviour over time is consistent across parameter space but quantitative variations are significant, as explained in Section \ref{sec:RedistribDiss}.

The fraction of heat dissipated over time does not vary much with $D$ but does vary with $A$. In Fig. \ref{fig:EventsOverTime} the fractional heat dissipated by time $t$ is approximately given by $(2/\pi) \arctan[(0.22 - 0.18A)(t/\tau - 2.3)]$, an arbitrary empirical fit. Tectonic activity persists longer, when $A$ is higher, because cell relaxation is less complete and a failed cell is more likely to fail for a second time, dissipating additional heat later in the star's life. Most tectonic activity occurs early in the star's life, with $50\%$ of the heat dissipated from $t/\tau \approx 7$ to $t/\tau \approx 19$ depending on $A$. This corresponds to a rotation frequency from $\approx 0.35\Omega(0)$ to $0.25\Omega(0)$. The star remains seismically active even after losing the majority of its rotational energy, because the strain since birth is stored in the crust in a metastable, far-from-equilibrium configuration, which relaxes via repeated failure on a time-scale longer than $\tau$. The run-to-run dispersion in energy released as a fraction of total energy released is small and is consistent across parameter space, with ${\rm var}(\Sigma \Delta E) \approx 10^{-4} \langle \Sigma \Delta E\rangle^2$.

\begin{figure}
\includegraphics[height=5.5cm,width=8.5cm]{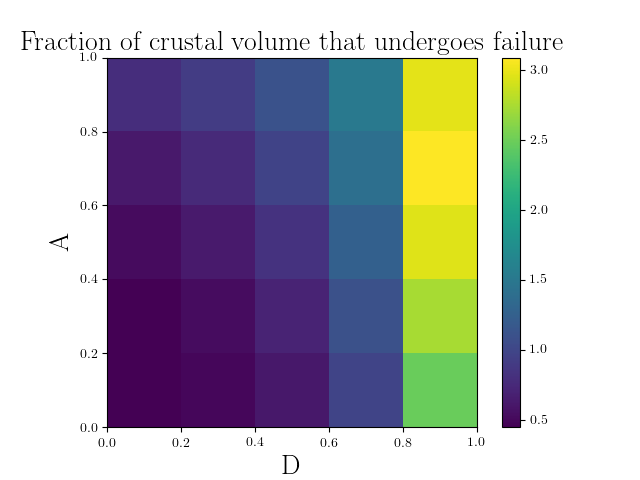}
\caption{The total fractional volume of the star's crust that fails vs $(A, D)$. This includes the same cell failing repeatedly; a given cell failing twice is equivalent to two cells (of the same size) failing once each, so the failed fraction can exceed 100\%. Dark blue corresponds to a small fraction of failed crust, $\approx 45\%$ and bright yellow corresponds to a large fraction, $\approx 300\%$. The data are from five simulations per value of $(A, D)$.}
\label{fig:AvEventNumber}
\end{figure}

Failure is a time-limited process; beyond a certain age, the star spins down sufficiently that the crust does not fail in any location during the remainder of its life. It is therefore meaningful to speak of the ultimate volumetric fraction of the crust that ever undergoes failure. The ultimate volumetric failure fraction is plotted versus $A$ and $D$ in Fig. \ref{fig:AvEventNumber}. The total volume of the crust that fails reaches a minimum near $(A, D) = (0, 0)$ with $\approx 45\%$ of the crust failing and a maximum near $(A, D) = (0.7, 0.9)$ with $\approx 300\%$ of the crust failing, where above $100\%$ means the same regions of the crust fail repeatedly. The volumetric failure fraction increases with $D$ because stronger nearest-neighbour interactions lead to larger avalanches. However the dependence on $A$ is not as straightforward. Cell relaxation is less complete, when $A$ is higher, through Eq. (\ref{eq:StrainUpdate}), which makes secondary failures more likely. On the other hand, the nearest-neighbour interaction is weaker, when $A$ is higher, which produces smaller avalanches. The effects in the previous two sentences counteract each other. If $D$ is small the nearest-neighbour interaction is always weak so the effect of $A$ on avalanche size is irrelevant.

\begin{figure}
\includegraphics[height=5.5cm,width=8.5cm]{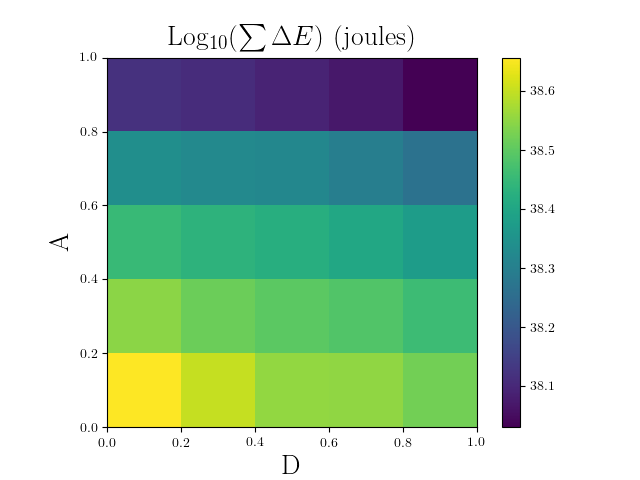}
\caption{Total heat dissipated (in joules) by all failure events in the star's life. Each data point is the average result of five simulations. Bright yellow corresponds to high values $\approx 5 \times 10^{38}$J, and dark blue to lower values $\approx 1 \times 10^{38}$J. The data are from five simulations per value of $(A, D)$.}
\label{fig:EnergyReleased}
\end{figure}

For $(A, D) = (0.5, 0.5)$ and $\sigma_{ i, j} = 0.1$ the corresponding energy density of failure is $\approx 2.7 \times 10^{26}$ Jm\textsuperscript{-3}. When $80\%$ of the crust fails the total heat released while moving the crust is $\approx 2.7 \times 10^{38}$ J. This is consistent with the results in Fig. \ref{fig:EnergyReleased} in which we display the total heat dissipated in all failure events for a variety of values of $(A, D)$.

\subsection{Redistribution and dissipation}
\label{sec:RedistribDiss}
The dissipation and redistribution of strain in the crust, governed by the parameters $A$ and $D$ respectively, have a significant impact on the quantitative outcomes of failure. In this section we examine the impact of the values of $A$ and $D$ on the volume of the crust that undergoes failure, the $\Delta E$ and $\Delta t$ PDFs, and the total heat dissipated while moving the crust in failure events.

\begin{figure}
\includegraphics[height=5.5cm,width=8.5cm]{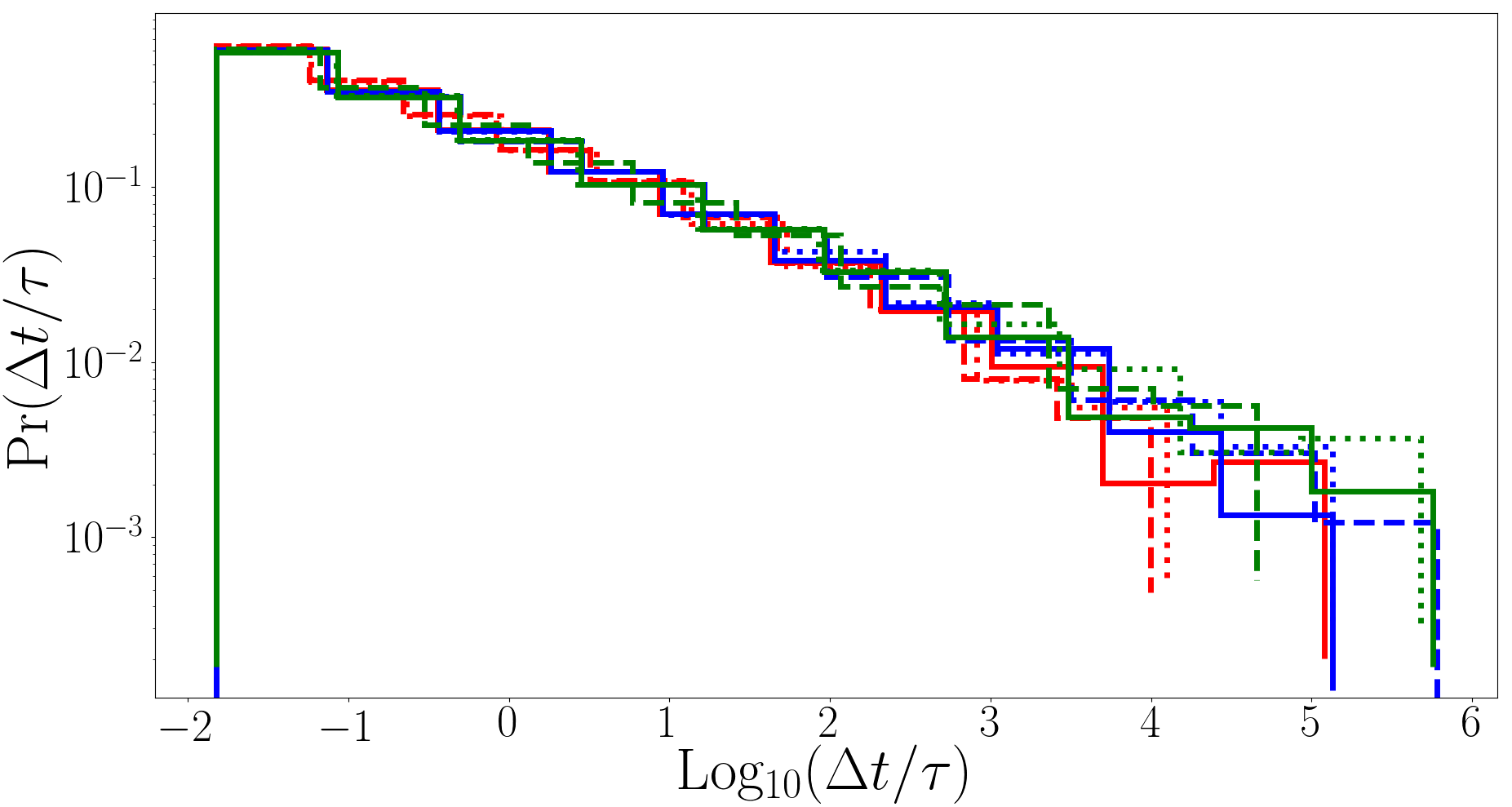}
\caption{Waiting-time PDFs for a range of values of $A$ and $D$. Red corresponds to $A=0.1$, blue to $A=0.5$ and green to $A=0.9$. The solid lines correspond to $D=0.1$, dashed to $D=0.5$ and dotted to $D=0.9$. Each PDF is constructed from all the events from five simulations.}
\label{fig:WaitHistogram}
\end{figure}

\begin{figure}
\includegraphics[height=5.5cm,width=8.5cm]{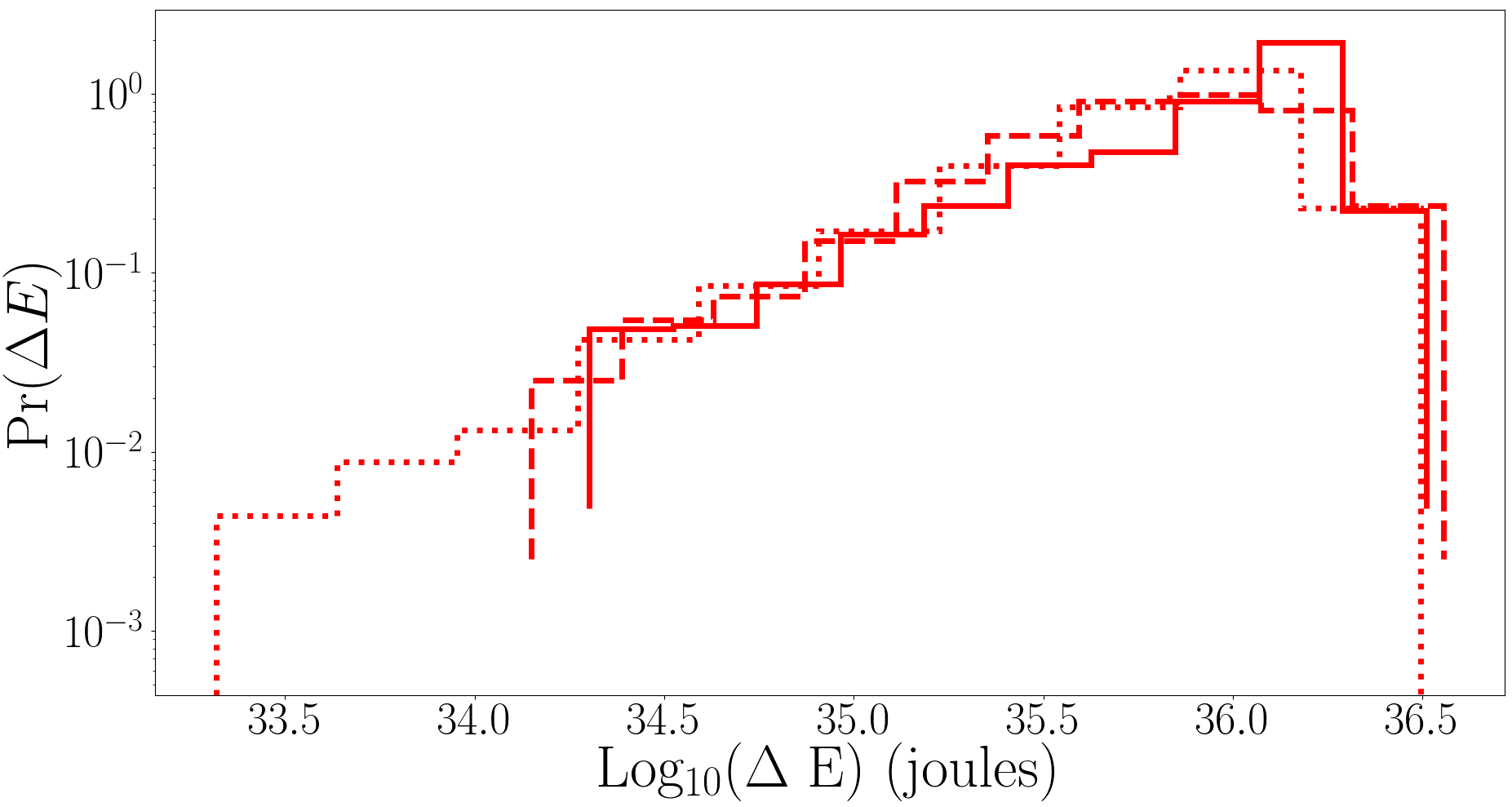}
\includegraphics[height=5.5cm,width=8.5cm]{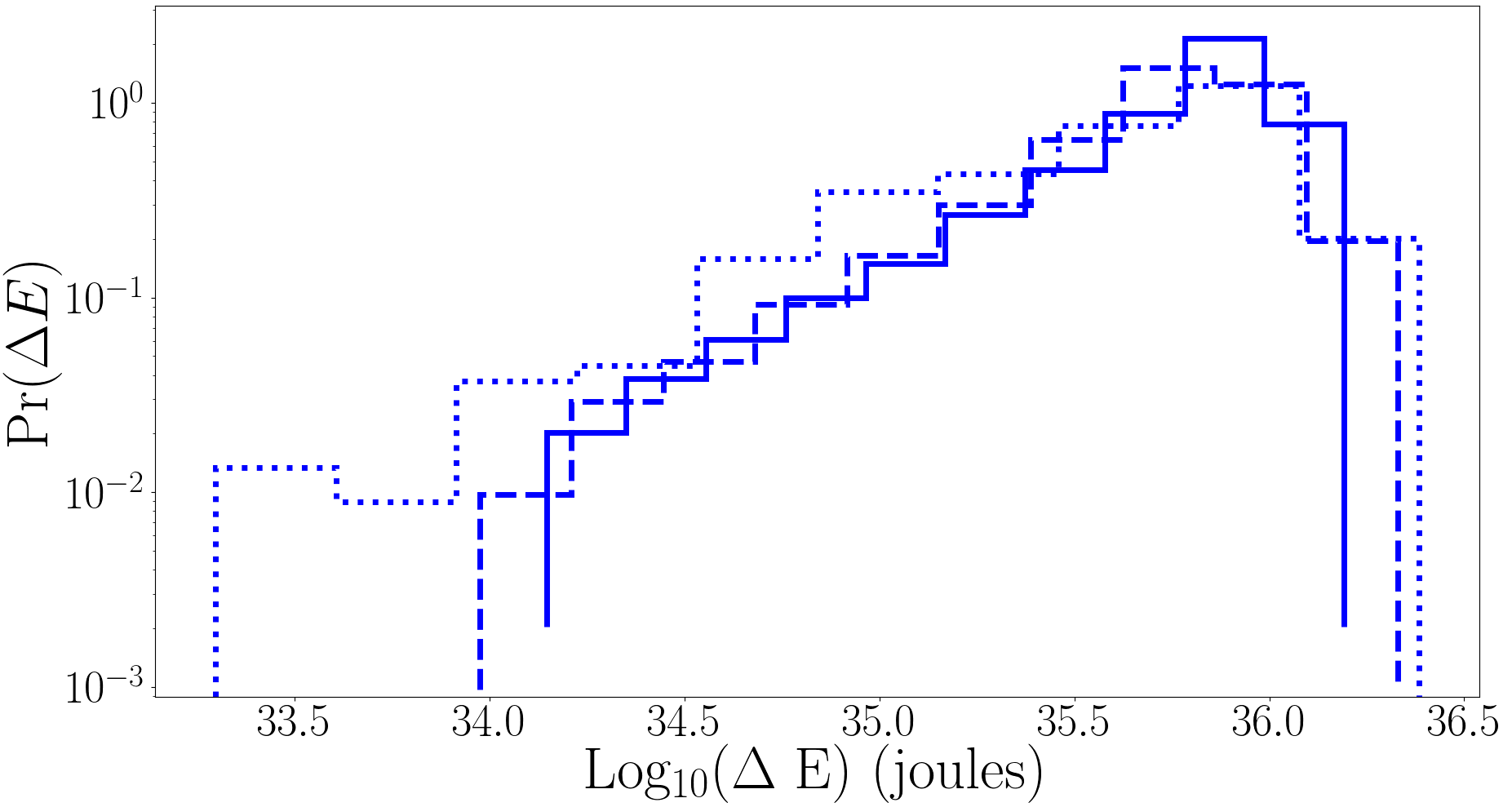}
\includegraphics[height=5.5cm,width=8.5cm]{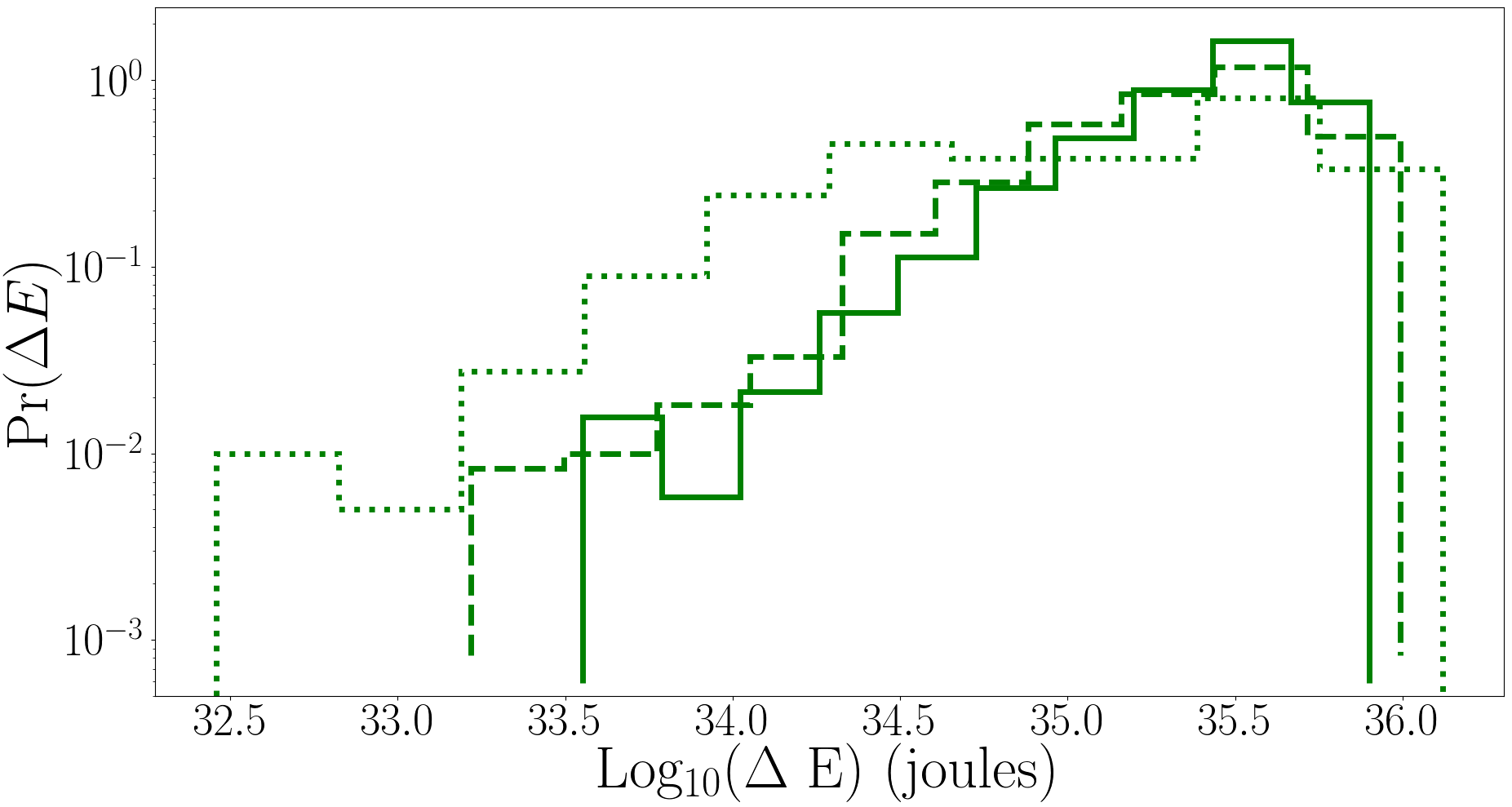}
\caption{Size PDFs for a range of values of $A$ and $D$. Red corresponds to $A=0.1$ (upper panel), blue to $A=0.5$ (middle panel) and green to $A=0.9$ (lower panel). The solid lines correspond to $D=0.1$, dashed to $D=0.5$ and dotted to $D=0.9$. Each PDF is constructed from all the events from five simulations. }
\label{fig:SizeHistogram}
\end{figure}

In Section \ref{sec:SizeWaitDistrib} we examine the PDFs of $\Delta E$ and $\Delta t$ for the representative values $(A,D)=(0.5,0.5)$. Now we turn to examining the effects of different values of $A$ and $D$ on the PDFs. In Fig. \ref{fig:WaitHistogram} the $\Delta t$ PDFs for a range of values of $A$ and $D$ are presented. Most immediately we can see that changing $A$ or $D$ has little to no effect on the $\Delta t$ PDF. This insensitivity to the parameters is because events are driven by the build up of strain due to spin down which is unaffected by $A$ or $D$. In contrast, the effects on the $\Delta E$ PDF, presented in Fig. \ref{fig:SizeHistogram}, are notable, with a shift in the locations of the peaks of approximately one order of magnitude from $(A,D)=(0.1, 0.1)$ to $(A,D)=(0.9,0.9)$. Events are smaller for larger $A$ and $D$. The dispersion of the PDF increases with $D$. Near $D=0.1$ the sizes are spread over approximately two orders of magnitude; near $D=0.9$ the sizes are spread over three orders of magnitude. The nearest-neighbour interaction is stronger with higher $D$ so avalanches are more common and larger and the dispersion is larger too.

Let us now return to $\Sigma \Delta E$, the total heat dissipated by failure. Fig. \ref{fig:EnergyReleased} yields a maximum $\approx 5\times10^{38}$J and a minimum $\approx 1\times10^{38}$ J. Unlike the volumetric failure fraction the total heat dissipated is strictly decreasing with $A$ and $D$. Larger $A$ and $D$ mean that events are smaller. Additionally, whereas the volumetric failure fraction is more sensitive to $D$ than to $A$, the reverse is true for $\Sigma \Delta E$; it is more sensitive to $A$ than to $D$. As $A$ increases $\Sigma \Delta E$ decreases significantly. Individual failed cells are less energetic and the nearest-neighbour interaction is also weaker. Not only are individual events smaller but avalanches are smaller and less frequent; both effects work to decrease $\Sigma \Delta E$. Conversely increasing $D$ strengthens the nearest-neighbour interaction causing avalanches to be larger and more frequent but individual cell failures to be less energetic; these effects partially counteract one another.
	
\section{Mass Ellipticity}
\label{sec:MassEllipticity}
As well as the statistics of failure events, we are also interested in the local mountains that failure events create. These mountains break axisymmetry, creating a non-zero ellipticity leading to continuous emission of gravitational waves which may be detectable by modern instruments such as LIGO-Virgo-KAGRA \citep{riles2013gravitational, woan2018evidence}. There have been several searches for continuous gravitational wave signals \citep{abbott2019narrow, abbott2019directional, abbott2019search,abbott2019searches1,abbott2019searches2, abbott2020gravitational, papa2020search}, but none have been found to date.

In this section the physical quantities of interest are the star's ellipticity and gravitational wave strain. These quantities are numerically convergent as a function of the angular velocity decrement, $\delta \Omega$, and the number of cells $N^2$ ; see Appendix D for further details.

\subsection{Definition}
\label{sec:Defn}
The moments of inertia of the star are necessary to calculate its ellipticity, $\epsilon$, and thus the characteristic wave strain, $h_0$, of the gravitational wave signal. The moment of inertia tensor, $I_{ij}$, is defined as
\begin{align}
I_{ij}=\sum_{k=0}^{N^2-1} m_{k}\Big(|\vec{r}^{(k)}|^2 \delta_{ij}-r^{(k)}_{i}r^{(k)}_{j}\Big).
\label{eq:MomOfIner}
\end{align}
In Eq. (\ref{eq:MomOfIner}) $N^2$ is the number of mass elements in the system, $m_k$ is the mass of the $k^{\rm th}$ element, $|\vec{r}^{(k)}|$ is the distance to the mass element from the origin (the centre of the star in this case) and $r^{(k)}_{i}$ is the $i^{\rm th}$ coordinate of the $k^{\rm th}$ mass element ($i = x,y,z$).

Local failure affects the moments of inertia of the star unequally, which breaks axisymmetry and leads to $\epsilon \neq 0$. In order to calculate the moments of inertia of the whole star we need the masses and centres of mass of each crustal cell and of the fluid core. The calculations of the foregoing quantities are detailed in Appendix E. As at the end of Section \ref{sec:CellAutomaton} we assume that the crust and core are of uniform density, with $\rho_{\rm crust} = 10^{17}$ kgm\textsuperscript{-3} \citep{horowitz2007phase}, and $\rho_{\rm core} = 6.38 \times 10^{17}$ kgm\textsuperscript{-3} and the crust is 1 km thick, i.e. $R-R'=1$ km. These fiducial values are selected for the sake of illustration; an exhaustive study of the parameter space is postponed to future work.

The off-diagonal moments of inertia are much smaller than the other moments of inertia; typically we find $|I_{xy}| \approx |I_{xz}| \approx |I_{yz}|\leq 10^{-9} |I_{zz}|$. The principal moments of inertia are well approximated by $I_{xx}$, $I_{yy}$, and $I_{zz}$. The ellipticity of interest for gravitational wave applications is given by
\begin{align}
\epsilon = \frac{\left|I_{xx}-I_{yy}\right|}{I_{zz}}. 
\label{eq:ellipticity}
\end{align} 
As the star is triaxial, there is a second ellipticity in the problem, namely $\epsilon_2 = |I_{xx}-I_{zz}|/I_{zz}$. This is dominated by the centrifugal bulge, which is axisymmetric and therefore does not emit gravitational waves. We therefore concentrate on $\epsilon$ in what follows.

\begin{figure}
\includegraphics[height=5.5cm,width=8.5cm]{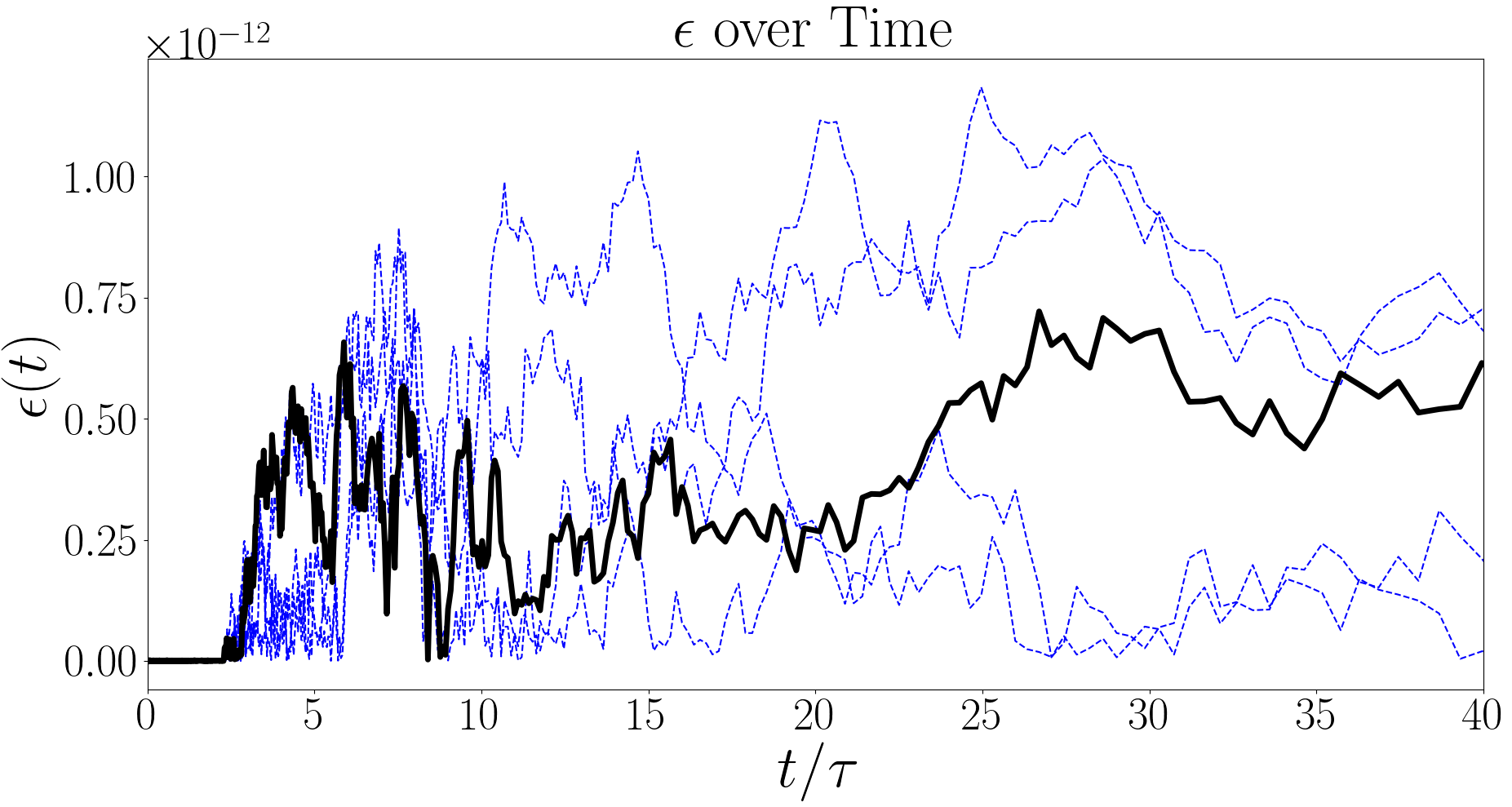}
\includegraphics[height=5.5cm,width=8.5cm]{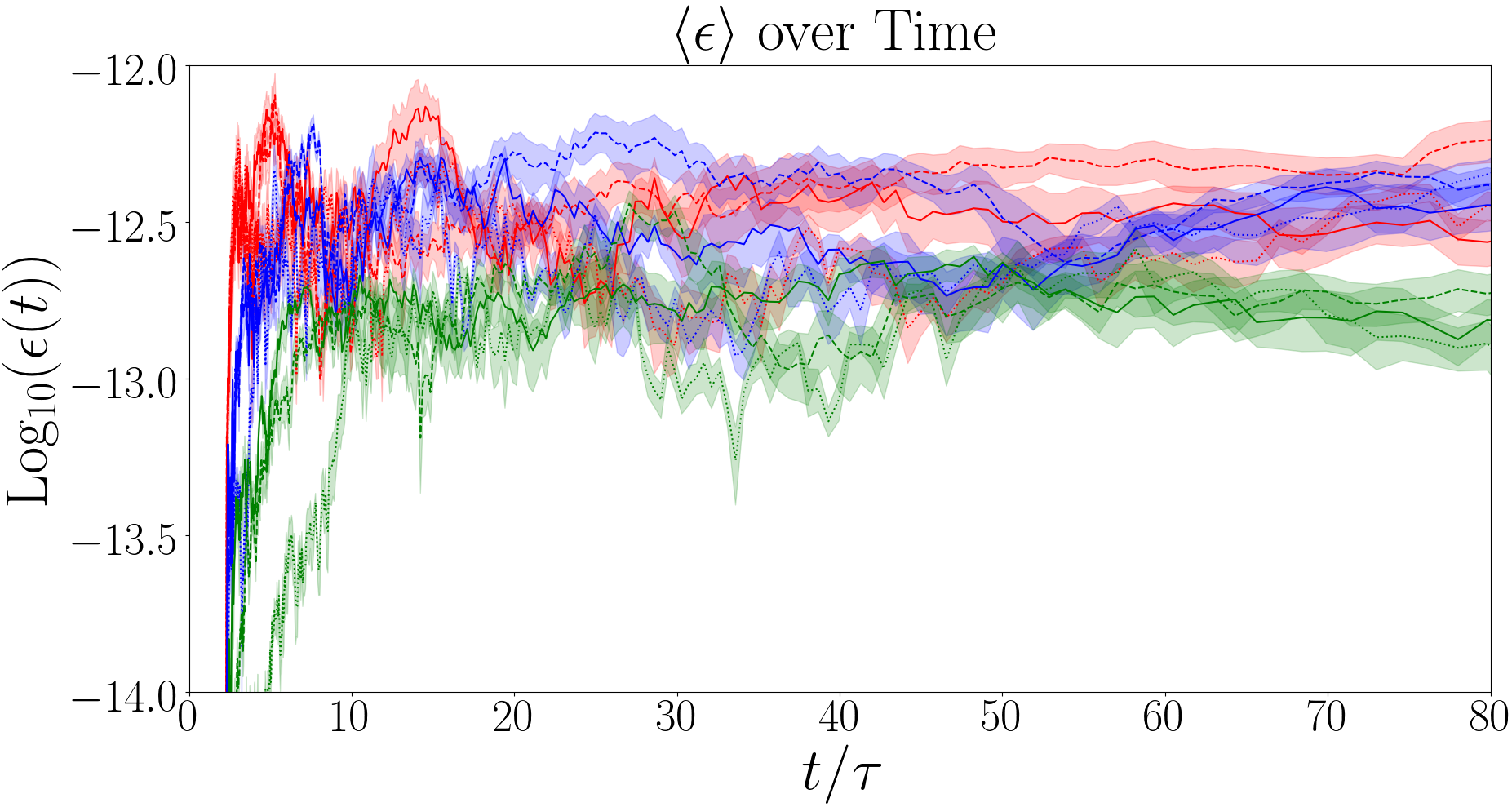}
\caption{Deformation as a function of time. (Top panel.) Ellipticity of five random realisations of the automaton for $(A,D)=(0.5,0.5)$. One run is highlighted in black for legibility. The other four are included to convey a sense of the dispersion in outcomes. (Bottom panel.) The logarithm of the average of five trials of the model for multiple values of $(A,D)$ with the shading being one quarter of a standard deviation. Red corresponds to $A=0.1$, blue to $A=0.5$ and green to $A=0.9$. The solid lines correspond to $D=0.1$, dashed to $D=0.5$, dotted to $D=0.9$.}
\label{fig:logevTime}
\end{figure}

\begin{figure}
\includegraphics[height=5.5cm,width=8.5cm]{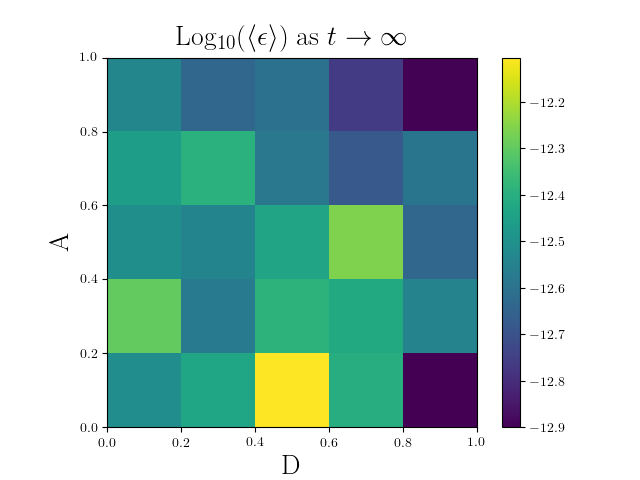}
\caption{The logarithm of the average ellipticity $\langle \epsilon \rangle$ as $t \rightarrow \infty$, where the average is taken over five trials. Bright yellow corresponds to high values $\approx 8 \times 10^{-13}$, and dark blue to lower values $\approx 1 \times 10^{-13}$. The data are from five runs per value of $(A,D)$.}
\label{fig:EllipticityStats}
\end{figure}

\subsection{Ellipticity evolution}
\label{sec:EllipEvo}
The long-term evolution of $\epsilon$ determines directly the long-term evolution of a tectonically active neutron star as a gravitational wave source. In Fig. \ref{fig:logevTime} we plot the ellipticity versus time for $0.1\leq A \leq 0.9$ and $0.1 \leq D \leq 0.9$. The differences between simulations of the same parameters are purely stochastic. The behaviour of $\epsilon$ is qualitatively similar across parameter space.

We obtain $\epsilon=0$ until the first failure at $t/\tau \approx 2.33 \pm 0.02$ then $\epsilon$ increases steeply. The behaviour is stochastic; individual events may increase or decrease $\epsilon$ but over the long term $\epsilon$ increases. As the rate of change of strain is independent of $A$ or $D$, the first failure always occurs at nearly the same time with small variations coming from $\sigma_{i,j}$. Fast spin down early in the star's life causes many failures, hence the fast rise in $\epsilon$. At late times $(t/\tau \gtrsim 200)$ failures happen less frequently and $\epsilon$ approaches a final value.

The standard deviation in $\epsilon$ is comparable to its mean, with ${\rm var}(\epsilon) \approx 0.25\langle \epsilon \rangle^2$ across parameter space, compared with the total heat dissipated, where we find ${\rm var}(\Sigma \Delta E)\approx 10^{-4}\langle \Sigma \Delta E\rangle^2$. This is because the locations of the failed cells, among other factors, determine the size and sign of the change in $\epsilon$. Suppose the first cell to fail is located at $\vec{r}_{i,j}=(r_{i,j},\pi/2,0)$, i.e. lying directly on the $x$-axis. As a result of this cell failing $I_{yy}$ increases but $I_{xx}$ does not\footnote{Strictly $I_{xx}$ does change due to the movement of the neighbouring cells, but the change is negligible for the purpose of this illustrative example.}. The first failure always increases $\epsilon$ because it breaks axisymmetry, but subsequent failures are not guaranteed to increase $\epsilon$. A second failure located at $\vec{r}_{i,j}=(r_{i,j},\pi/2,\pi/2)$, i.e. on the $y$-axis, increases $I_{xx}$ but not $I_{yy}$, causing $\epsilon$ to decrease. However if instead the second failure is located at $\vec{r}_{i,j}=(r_{i,j},\pi/2,\pi)$, i.e. on the negative $x$-axis, $\epsilon$ increases further. Failure events increase or decrease $\epsilon$ by various amounts depending on where the failed cells are located relative to previous failures. The change in $\epsilon$ also depends on factors such as the size of $\Delta r_{i,j}$ and the number of cells that fail, but the spatial location of failures is the most significant source of dispersion.

In Fig. \ref{fig:EllipticityStats} we present the final ellipticity of the star, as it varies with $A$ and $D$. The average final $\epsilon$ varies between $\approx 1 \times 10^{-13}$ and $\approx 8 \times 10^{-13}$ across parameter space. However the dispersion is large and individual simulations can achieve ellipticities over $10^{-12}$. It is apparent the final ellipticity of the star is insensitive to $A$ and $D$. As $D$ is increased, $\Delta r_{i,j}$ becomes smaller, through Eq. (\ref{eq:FailEnergy}), and the nearest-neighbour interaction becomes stronger. Cells move less upon failure, causing $\epsilon$ to decrease, but more cells fail, causing $\epsilon$ to increase. These two effects counteract one another meaning a weak net effect on $\epsilon$. As $A$ is increased, $\Delta r_{i,j}$ becomes smaller and the nearest-neighbour interaction becomes weaker, both of which cause $\epsilon$ to decrease. But larger $A$ allows cells to fail multiple times causing $\epsilon$ to increase. Again these effects counteract one another meaning a weak net effect on $\epsilon$.

Although the final ellipticity of the star is insensitive to $A$ and $D$ this is not the case at early times. In Fig. \ref{fig:logevTime} we see that the ellipticity is larger early in the star's life for smaller values of $A$. As discussed in Sections \ref{sec:SizeWaitCorr} and \ref{sec:History} and seen in Fig. \ref{fig:EventsOverTime} tectonic activity is more evenly spread over the star's life for large $A$. When $A$ is small, $\epsilon$ approaches its final value quickly as tectonic activity is relatively high, outpacing simulations with large values of $A$. The simulations with large and small $A$ approach similar $\epsilon$ values as the star ages. In addition we find that $\epsilon$ increases, when $\beta$ decreases, as expected from Eqs. (\ref{eq:FailEnergy}) and (\ref{eq:NeighbourFailEnergy}). The failure induced crustal movements are larger when $\beta$ is smaller.

\subsection{Gravitational wave strain}
\label{sec:Wavestrain}
\begin{figure}
\includegraphics[height=5.5cm,width=8.5cm]{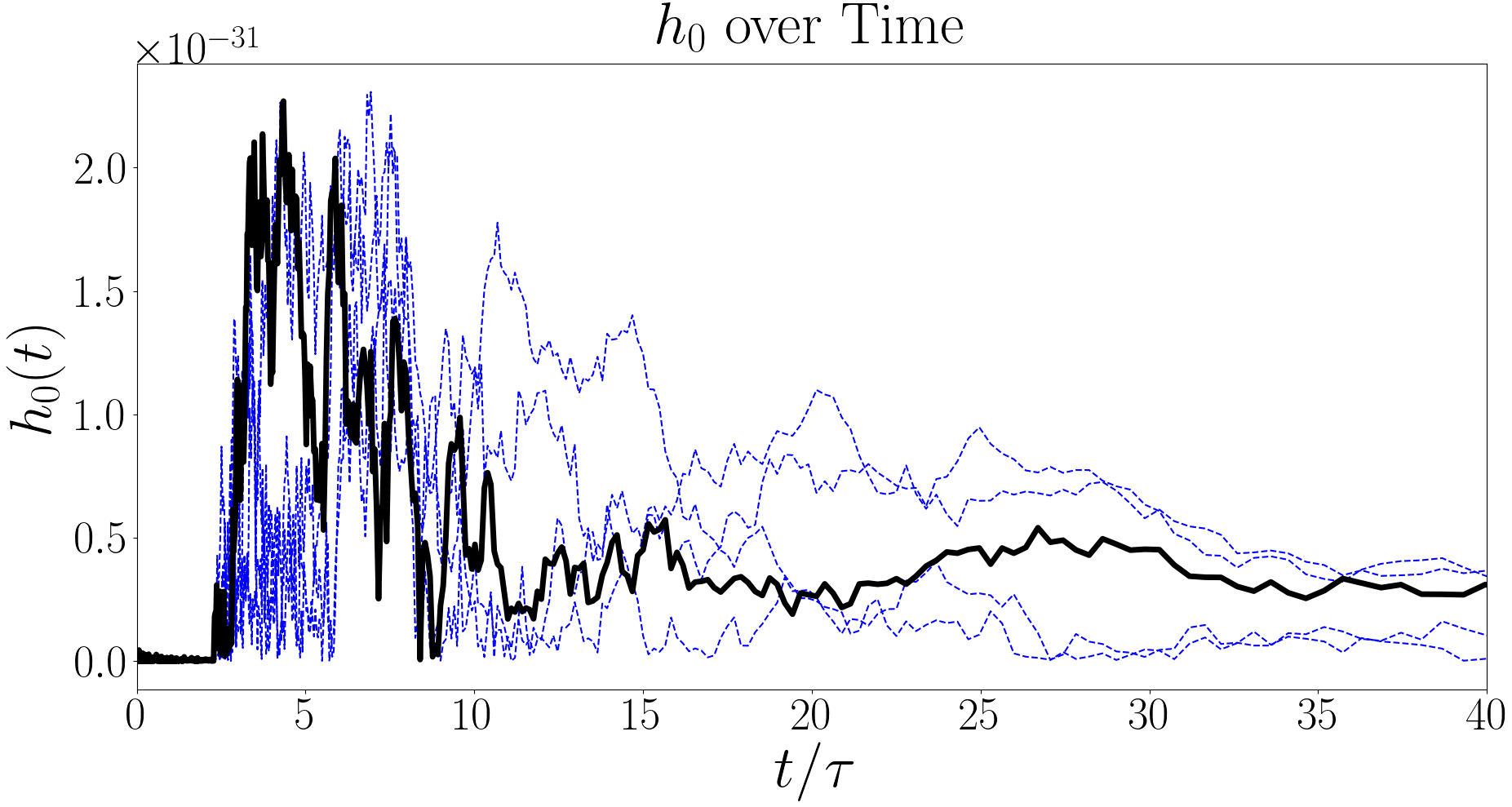}
\includegraphics[height=5.5cm,width=8.5cm]{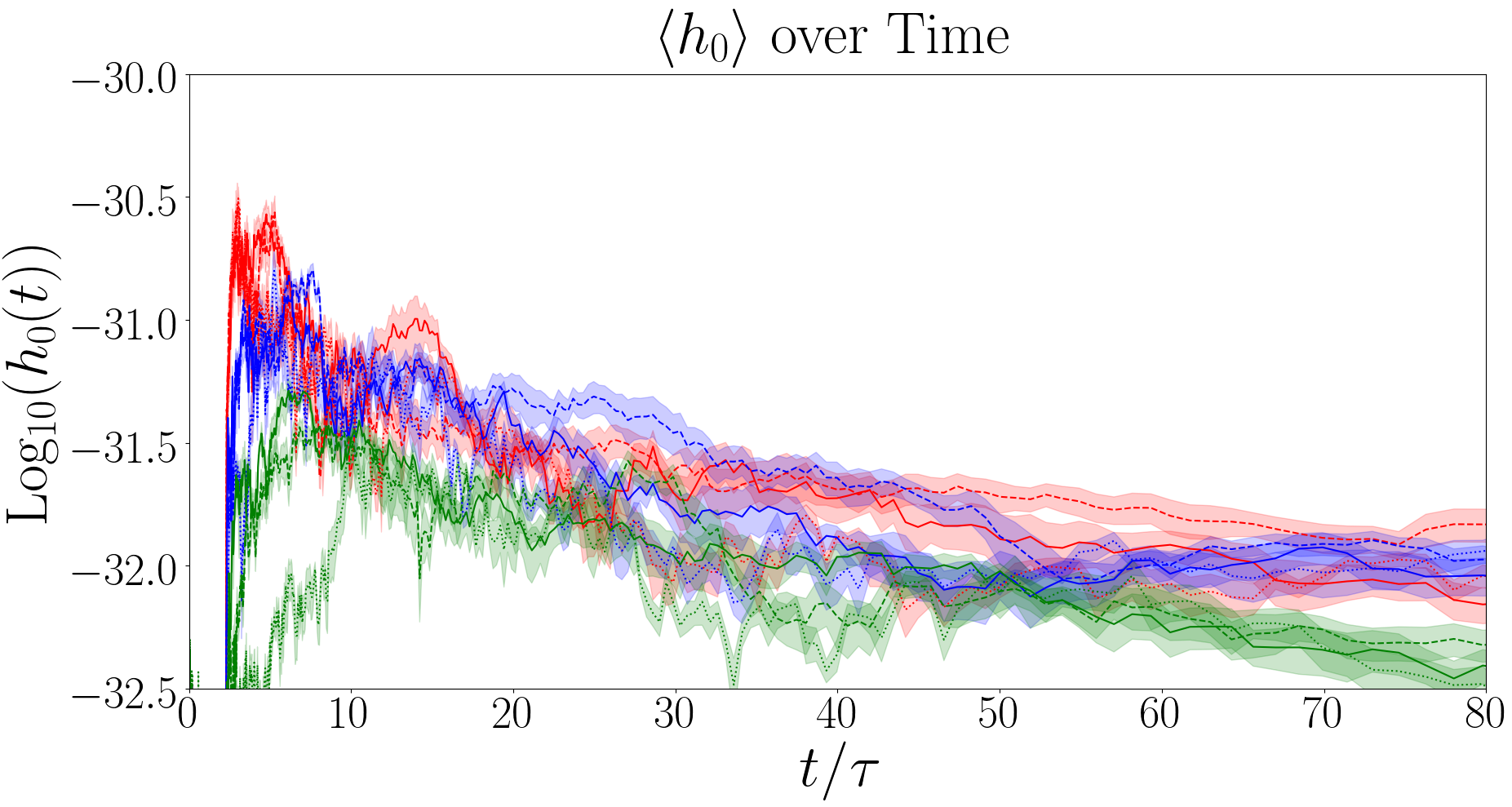}
\caption{Gravitational wave signal as a function of time. (Top panel.) Wave strain of five random realisations of the automaton for $(A,D)=(0.5,0.5)$. One run is highlighted in black for legibility. (Bottom panel.) The logarithm of the average $h_{0} $ of each set of five simulations with the shading being one quarter of a standard deviation. Red corresponds to $A=0.1$, blue to $A=0.5$ and green to $A=0.9$. The solid lines correspond to $D=0.1$, dashed to $D=0.5$, dotted to $D=0.9$. The upper and lower panels are constructed from the same respective data sets as the upper and lower panels of Fig. \ref{fig:logevTime}.}
\label{fig:loghvTime}
\end{figure}
Any rotor that is not symmetric about its rotation axis emits gravitational waves. The gravitational wave strain of a star with ellipticity $\epsilon$, rotating at frequency $f$ about the $z$-axis and at distance $d$ from the Earth, is given by \citep{jaranowski1998data, horowitz2010gravitational},
\begin{align}
h_{0}=&\frac{16\pi^2G}{c^4}\frac{\epsilon I_{zz} f^2}{d},
\label{eq:wavestrain}\\
=&3.8\times10^{-31}\left(\frac{d}{1 \rm kpc}\right)^{-1}\left(\frac{\epsilon}{10^{-12}}\right)\left(\frac{I_{zz}}{10^{38} \rm kg m^{2}}\right) \left( \frac{f}{300 \rm Hz} \right)^2.
\label{eq:fiducial}
\end{align}
Eq. (\ref{eq:fiducial}) is evaluated from Eq. (\ref{eq:wavestrain}) for fiducial astrophysical parameters. Unless otherwise noted we calculate the wave strain for an object that is $1$ kpc away from Earth in what follows.

In Fig. \ref{fig:loghvTime} we present the wave strain of the gravitational wave signal from the star for a variety of values of $(A,D)$. In the upper panel five randomly chosen simulations are presented for $(A,D)=(0.5,0.5)$. In the lower panel we present the average wave strain for a variety of values of $(A,D)$ on a logarithmic scale. The shading indicates a one quarter standard deviation spread. As in Eq. (\ref{eq:wavestrain}), with $h_0 \propto \epsilon \Omega^2$, there is a steep rise in $h_0$ to a maximum in the range $h_0 \lesssim 10^{-30}$ for $4 \lesssim t/\tau \lesssim 15$, corresponding to $0.25 \lesssim \Omega(t)/\Omega(0) \lesssim 0.45$. A gradual decline follows, as $\epsilon$ approaches a constant value, and $\Omega(t) \propto (1 + t/\tau)^{-1/2}$ decreases for increasing $t$.

The two dynamical effects that determine $h_0$ are the changing geometry of the star and the decaying rotational frequency. As explained above, individual events may increase or decrease $\epsilon$, which in turn affects $h_0 \propto \epsilon \Omega^2$. During the sharp initial rise in $\epsilon$, seen in Figs. \ref{fig:logevTime} and \ref{fig:loghvTime}, the increase in $\epsilon$ outpaces the decrease in $\Omega(t)^2$. However as spin down continues the rate of failure slows and the frequency decay becomes the dominant factor causing $h_0$ to tend towards zero. As with $\epsilon$ the variation in $h_0$ is also large with ${\rm var}(h_0) \approx 0.25\langle h_{0}\rangle^2$ across parameter space.

As the wave strain is proportional to $\epsilon$ it has the same behaviour with $A$ and $D$. In Fig. \ref{fig:loghvTime} the peak $h_{0}$ in a given run reaches a maximum $\approx 5 \times 10^{-31}$ near $A=0$, and decreases to a minimum $\approx 5 \times 10^{-32}$ near $A=1$. However the dispersion is large and individual simulations can achieve wave strains close to $10^{-30}$.

The predicted wave strain in Eqs. (\ref{eq:wavestrain}) and (\ref{eq:fiducial}) is somewhat too low to be detected by current generation detectors \citep{aasi2015advanced}, though there is room to increase it by increasing $f$ for example. However it is within reach of next generation detectors. The signal scales in proportion to $\epsilon$ and therefore $\Delta r_{i,j}$, $\Delta E$, and $\mu$ as well as the scalings with $f$ , $d$, and $I_{zz}$ in Eq. (\ref{eq:wavestrain}). Importantly, it also depends on the cell volume, which in turn depends on the characteristic dimension of material imperfections like grain boundaries (see Section \ref{sec:Micro}). It is challenging to predict from first principles the dimension of such mesoscopic features in terrestrial materials, let alone in neutron star matter, where there are no controlled lab- oratory experiments as a guide. Finally, $h_0$ depends on $\beta$. In this paper, we take $\beta=0.9$, an astrophysically conservative assumption; smaller values of $\beta$ are plausible and lead to higher $h_0$.

The wave strain reaches a maximum in the range $\Omega(t) \approx 0.25\Omega(0)$ to $0.45\Omega(0)$, i.e. when the star spins at 200 to 350 Hz $(t/\tau \approx 4$ to $15)$. Using the value of $\tau$ from the end of Section \ref{sec:CellAutomaton} as one possible illustrative example, a pulsar with a birth frequency of 800Hz would reach its maximum wave strain $5 \times 10^3$ years to $2 \times 10^4$ years after its birth. The wave strain peaks later for larger $A$ because, as discussed in Section \ref{sec:EllipEvo} and seen in Fig. \ref{fig:logevTime}, $\epsilon$ is smaller at early times for larger $A$ and reaches a final value more slowly.

\section{Conclusions}
\label{sec:Conclusion}

In this paper we investigate the long-term, far-from-equilibrium behaviour of repeated, macroscopic, failure events in a neutron star crust. The resulting tectonic process forms mountains which emit gravitational waves. We develop an idealised cellular automaton to represent the crust, where individual cells are small finite crust elements whose location evolves over the course of spin down. We make use of a critical strain threshold in each cell and a nearest-neighbour interaction to model the strain redistribution and thermal dissipation of mechanical failure. The global build up of strain is driven by spin-down deformation and modelled using the formalism developed by \cite{franco2000quaking}. While \cite{horowitz2009breaking} found that neutron star crust material appears to fail in a global sense rather than cracking locally due to the extreme pressure, the molecular dynamics simulations are on the scale of $\sim 10^{11}$ cubic femtometres of material. The automaton in this paper involves cells on the scale of $\approx 1 $km, where mesoscopic and macroscopic inhomogeneities (e.g. grain boundaries and seismic faults) may plausibly become significant.

We find a correlation between the size of an event and the time until the next event, with the size and direction of the correlation depending on $A$. Correlations are strong and positive for $A=0.9$, with Spearman rank coefficient $\approx 0.3$ but are strong and negative for $A=0.1$ with Spearman rank coefficient $\approx -0.6$. This behaviour is due to waiting times lengthening and the volume of equatorial cells increasing as the star ages, and small values of $A$ suppressing avalanches for late events.

We find a power-law PDF for the waiting-times between events which is unaffected by $A$ or $D$. The qualitative form of the size ($\Delta E$) PDF is similar across parameter space. It is singly peaked and extends over approximately two orders of magnitude near $D=0$ with the dispersion growing to approximately three orders of magnitude near $D=1$. The increase in dispersion comes from an increased propensity for avalanches.

The average total heat dissipated while moving the crust varies
from $\approx 1 \times 10^{38}$J to $\approx 5 \times 10^{38}$J, reaching a minimum near $(A,D)=(1,1)$ and maximum near $(A,D)=(0,0)$. We find that the star dissipates heat faster for smaller values of $A$ with no dependence on $D$. A star with $A=0.1$ dissipates half of its total heat by $t \approx 7\tau$ whereas a star with $A=0.9$ dissipates half of its heat by $t \approx 19\tau$. The halfway points correspond to $\Omega(t) \approx 0.35\Omega(0)$ and $\Omega(t) \approx 0.25\Omega(0)$ respectively. Most of the energy is released early in the star's life, but activity continues beyond that point.

The key astrophysical implications of the paper are (1) the star needs to be born spinning at $\gtrsim 750$Hz in order for crustal failure to occur at all, consistent with the findings of \citet{fattoyev2018crust}; and (2) once the initial strain is lodged in the system, tectonic activity persists until $\Omega(t)/2\pi \approx 0.01\Omega(0)/2\pi \ll 750$Hz. This slightly counter-intuitive result has a straightforward explanation. Failure-driven avalanches redistribute strain widely throughout the star; they propagate until every cell inside the avalanche perimeter is subcritical. When the number of cells is large ($4\times 10^4$ in this paper), it is probable that some subcritical cells end up just below the threshold following an avalanche, i.e.\ $\sigma_{i,j} - \gamma_{i,j} \ll \sigma_{i,j}$ for some $(i,j)$. Such nearly critical cells are primed to fail, even when the incremental strain added by spin down is small at late times, e.g.\ for $\Omega(t) \approx 0.01\Omega(0)$. In general for $0<A<1$ and $0<D<1$ failure redistributes energy from one cell to the next in the manner described in Section \ref{sec:Macro} triggering more failures even when one has $\Omega(t)/2\pi < 750$Hz. The foregoing result does not rule out that crustal failure plays a role in glitches, but it means it is unlikely to be the dominant factor at play in all objects at all times. It has been suggested that failure events could leave cracks or other long-term defects that interact with the vortices in the superfluid core \citep{middleditch2006predicting}. The residual strain lodged in a cell after it fails for the last time is never lost.

We find an average final ellipticity between $\approx 1 \times 10^{-13}$ and $\approx 8 \times 10^{-13}$ across parameter space, but the dispersion is large and ellipticities between $5 \times 10^{-14}$ and $1 \times 10^{-12}$ occur routinely. The ellipticity depends on the cell volume, which is hard to predict theoretically, because it depends on the dimension of mesoscopic and macroscopic imperfections like grain boundaries and seismic faults, which result from hysteretic, far-from-equilibrium processes. The associated wave strain peaks at $h_0 \lesssim 10^{-30}$ for $\Omega(t) \approx 0.25\Omega(0)$ to $\approx 0.35\Omega(0)$ i.e. $\approx 200$Hz to $\approx 350$Hz. The wave strain peaks later for larger values of $A$ but is insensitive to $D$. The peak wave strain reaches a maximum near $A=0$ and a minimum near $A=1$. Of course, mountain formation due to spin-down deformation is not the only proposed mechanism for a continuous gravitational wave source. Accreting neutron stars are potential sources \citep{ushomirsky2000deformations} as are low-mass X-ray binary systems \citep{ushomirsky2000gravitational} and r-mode oscillations \citep{andersson2001r}.

In this paper we take $\beta=0.9$ for the phenomenological dissipation coefficient in the model. In other words during failure $90\%$ of the work done deforming the crust is converted into heat due to the deformation being plastic not elastic. The choice $\beta=0.9$ is guided by historical convention inspired by terrestrial materials in the absence of a detailed analysis of the neutron star crust, which is challenging to do reliably in the absence of controlled laboratory experiments on bulk nuclear matter. The choice is deliberately conservative. By reducing $\beta$, one can increase the amplitude of observable effects by increasing $\Delta r_{i,j}$, $\epsilon(t)$, and $h_{0}(t)$.

We plan to improve this calculation in the future. We will incorporate the improved description of strain build up in the absence of failure introduced by \cite{giliberti2019incompressible} and \cite{giliberti2020modelling} to extend the schema of \cite{franco2000quaking}. These extensions account for stratification of the star, e.g. continuous or discrete (crust and core), and model perturbations in the composition away from chemical equilibrium. \cite{giliberti2021starquakes} also considered accreting systems. It will be interesting to apply the tectonic process investigated in this paper to accreting systems as well.

\section*{Acknowledgements}

We thank the anonymous referee for their insightful comments and their substantial and concrete suggestions about how to improve the automaton. Specifically the referee contributed: (i) the improved energy conserving nearest-neighbour interaction in Section \ref{sec:Macro}; (ii) the associated rule in Section \ref{sec:RadialMovement} for neighbouring cells to rise and fall in response to failure, which conserves the average centrifugal-gravitational potential energy of the crust; (iii) a clearer description of the interface between the superfluid core and rigid crust; (iv) the inclusion of irreversible heat release as part of the failure process, as discussed in Sections \ref{sec:Macro} and \ref{sec:ThermalLosses} and Appendix B; and (v) the decision to evaluate $(r_{i,j},\theta_{i,j},\phi_{i,j})$ at the crust-core boundary.

This research was supported through the Australian Research Council Centre of Excellence for Gravitational Wave Discovery (grant number CE170100004) and Discovery Project DP170103625.

A. D. Kerin is supported by an Australian Government Research Training Program Scholarship and by the University of Melbourne.

\section*{Data Availability}
The data and code underpinning this article are available in the article and will be shared upon request to A. D. Kerin.



\bibliographystyle{mnras}
\bibliography{NSC-Bib}


\section*{APPENDIX A: DEFINITION OF CELL VOLUMES}

\begin{figure}
\includegraphics[height=5.5cm,width=8.5cm]{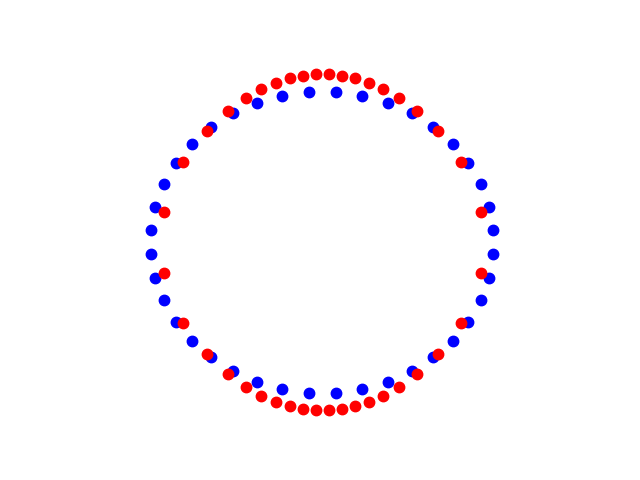}
\caption{A meridional cross-section of the star at $t = 0$ (blue dots) and $t \rightarrow\infty$ (red dots). The blue dots are located at $R=10.5$km with $e=0.1$ and are evenly spaced in $\theta$. The red dots define the star's final configuration with the blue dots as the initial points. The star's final configuration is slightly non-spherical due to small inaccuracies introduced by approximations when calculating the deformation vector $\vec{u}$. In the model of \citet{franco2000quaking}, $\vec{u}$ is calculated as a perturbation relative to a spherical background, even when the background is significantly ellipsoidal for $\Omega(t) / 2\pi \gtrsim 0.5 \, {\rm kHz}$. For example in \citet{franco2000quaking} the zero shear and pressure continuity boundary conditions [Eqs. (\ref{eq:defCoeffs1})--(\ref{eq:defCoeffs4})] are evaluated at the non-rotating radii $R'$ and $R$, rather than the true, ellipsoidal crust-core and crust-vacuum surfaces. The only term that explicitly acknowledges the non-spherical background is the centrifugal force [Eq. (15) in \citet{franco2000quaking}]. The spherical approximation means that the deformation vectors are accurate near the end of spin down, but small errors occur early on which contribute to the cumulative deformation at late times depicted by the red dots in the figure. Additionally $\vec{u}$ at the poles is smaller than $\vec{u}$ at mid-latitudes, as seen in Fig. 2 of \citet{franco2000quaking}. Indeed, for highly eccentric initial configurations with $e\gtrsim 0.9$, a small ``depression'' or ``valley'' forms at the poles.}
\label{fig:PolarSpread}
\end{figure}

\begin{figure}
\includegraphics[height=5.5cm,width=8.5cm]{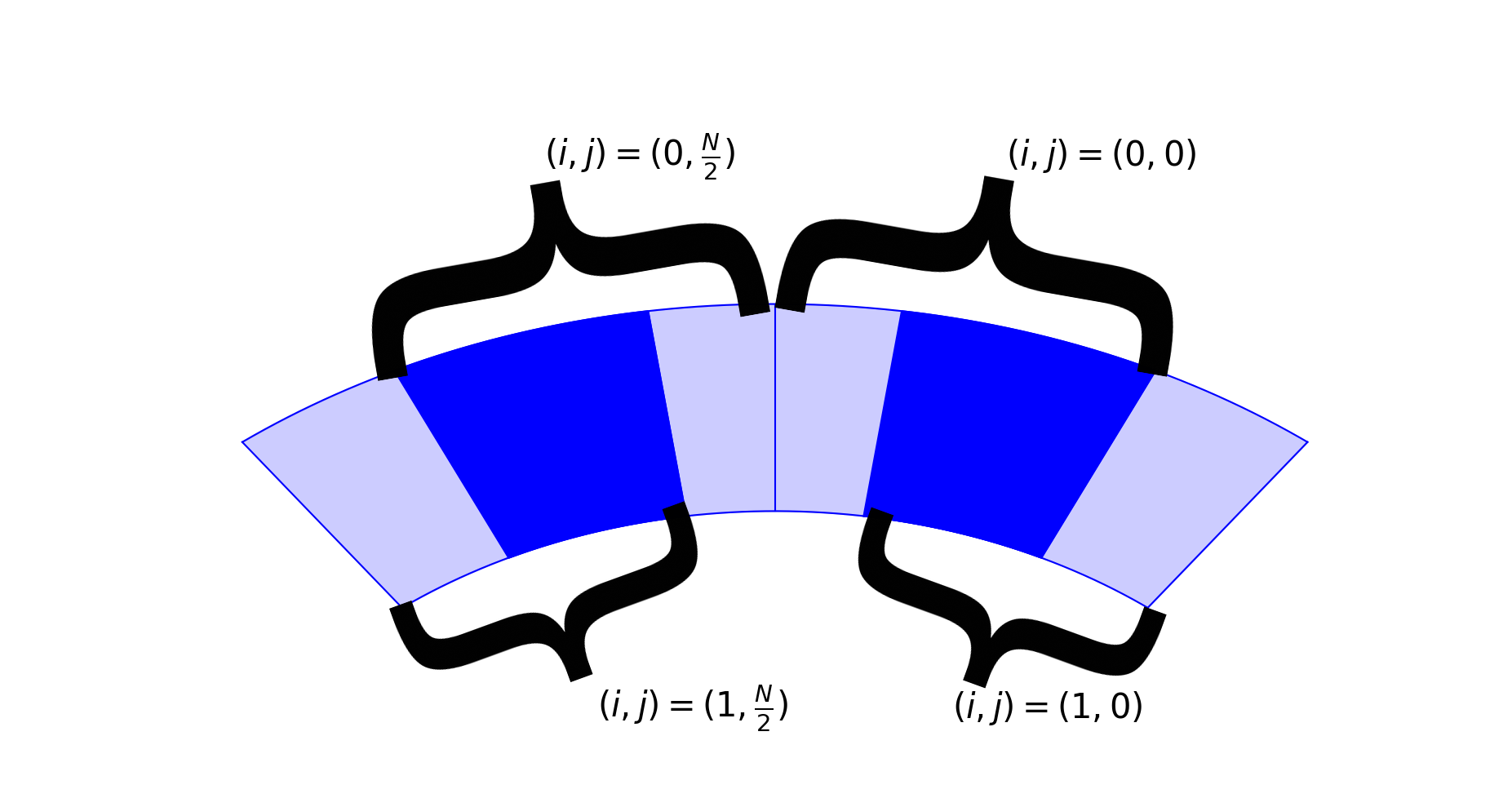}
\includegraphics[height=5.5cm,width=8.5cm]{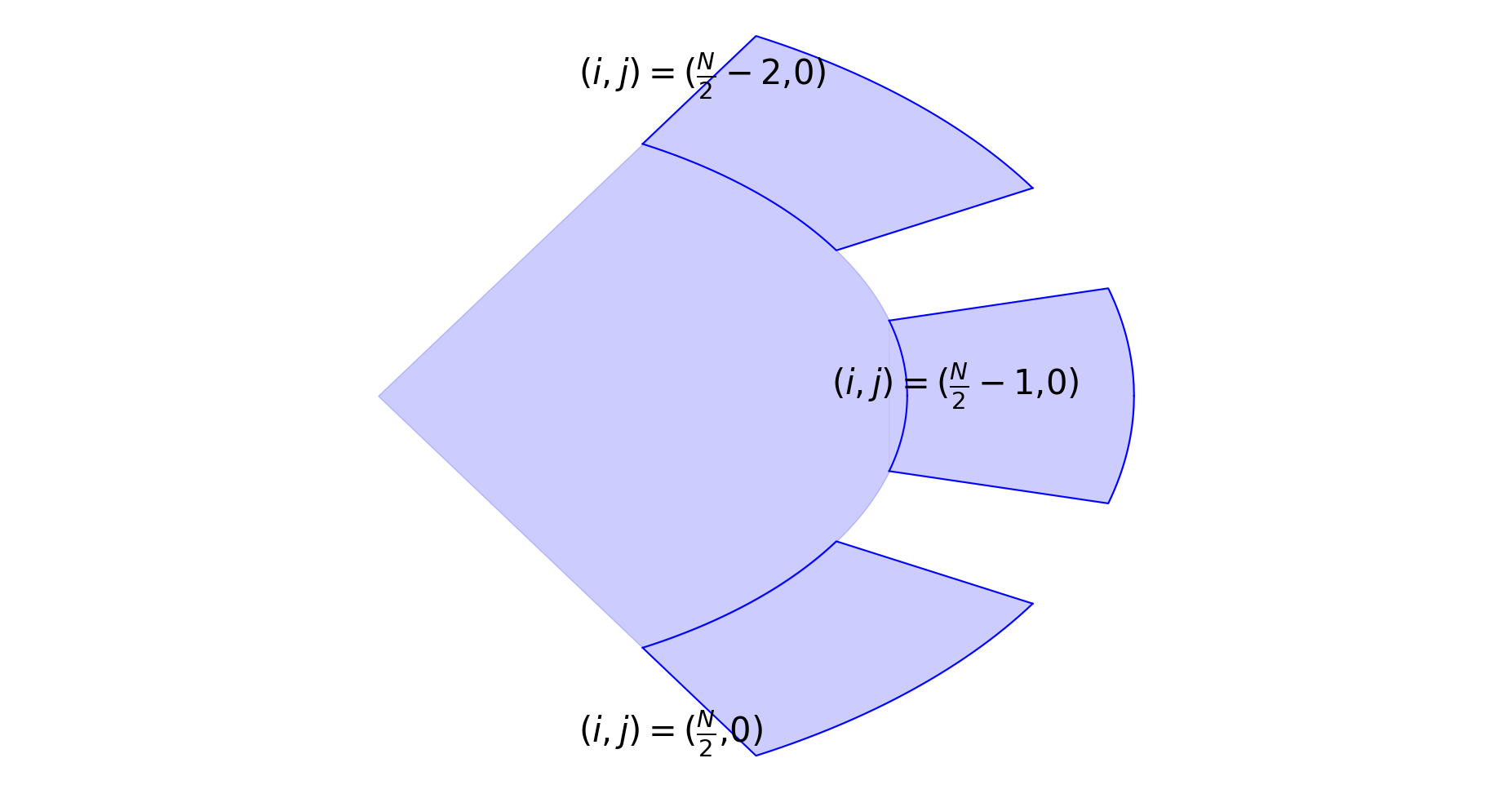}
\caption{The overlap of cells at the poles and gaps between cells at the equator in the hypothetical case where cells have constant solid angle, as detailed in Appendix A. The above configuration is avoided in this paper in favour of the configuration in Fig. \ref{fig:CellVolumeDiagram}. (Top panel.) A meridional cross-section near the north pole. The braces indicate distinct cells. The darker regions are regions where two cells overlap and the lighter regions are areas of no overlap. (Bottom panel.) A meridional cross-section near the equator. The three cells near the equator form gaps that expose the fluid core, which is unphysical. Hence the approach in Fig. \ref{fig:CellVolumeDiagram} is preferable.}
\label{fig:Overlap}
\end{figure}
In Section \ref{sec:Macro} we define the volume of a cell with indices $(i,j)$ and coordinates $(r_{i,j},\theta_{i,j},\phi_{i,j})$ as $r_{i,j} \leq r \leq r_{i,j}+(R-R')$, $\arccos\{[\cos(\theta_{i-1,j})+\cos(\theta_{i,j})]/2\}< \theta <\arccos\{[\cos(\theta_{i,j})+\cos(\theta_{i+1,j})]/2\}$, $(\phi_{i,j-1}+\phi_{i,j})/2 < \phi < (\phi_{i,j}+\phi_{i,j+1})/2$. In Fig. \ref{fig:CellVolumeDiagram} we see top-down and side-on cross-sectional views of the cell volumes. In this model cells have a constant depth $R-R'$ as we assume that the crustal thickness is a constant. This definition of cell boundaries along with Eqs. (\ref{eq:Initialise1}), (\ref{eq:Initialise2}) and (\ref{eq:Initialise3}) means that all cells are of equal volume at the time of initialisation.

We use the formalism of \citet{franco2000quaking} to model the secular deformation of the crust as the star spins down. In this schema the movement of the volume elements of the crust is not purely radial; there is a polar component, i.e. $u_{\theta}(\vec{r}) \neq 0$. Due to the polar movement the cells tend to cluster at the poles and spread out from the equator. This can be seen in Fig. \ref{fig:PolarSpread} where the initial pre-spin-down (blue dots) and final post-spin-down configurations (red dots) of the crust are presented; see also the second figure of \citet{franco2000quaking} where the deformation vectors at the crust are superposed on a meridional cross-section of the star. If we were to define cells to have constant solid angles then cells would overlap as they cluster about the poles and leave gaps as they spread out from the equator, as demonstrated in Fig. \ref{fig:Overlap}. Such an overlap is unphysical; the fluid core is covered by the crust completely at the equator (and everywhere else) and ``wrinkles'' do not form at the poles. To avoid this we define the polar boundaries between cells to lie halfway between them: a given cell $(i,j)$ occupies a polar region given by $\arccos\{[\cos(\theta_{i-1,j})+\cos(\theta_{i,j})]/2\}< \theta <\arccos\{[\cos(\theta_{i,j})+\cos(\theta_{i+1,j})]/2\}$. The location of the boundary is updated at every time-step as $\theta_{i,j}$ evolves with spin down.

The updating of boundaries causes the volumes of cells to change, as implied by Fig. \ref{fig:PolarSpread}. Equatorial cells become larger and polar cells become smaller, respectively increasing and decreasing the total elastic energy in those cells. As a result the total automaton-wide elastic energy is not conserved in general. Specifically equatorial regions accumulate strain more quickly than polar regions, so updating the cell boundaries artificially adds additional energy to the system. We check that this effect is modest by running the automaton with cell movement disabled (and also without failure for clarity of interpretation). We find that the total mechanical potential energy of the crust reaches $\approx 8.5\times10^{38}$J when the star is totally spun down. This is $\approx26\%$ less than an automaton where failure is disabled but cell movement is enabled. Proportionally the $26\%$ variation is significantly less than the variation in $\Sigma \Delta E$ from uncertain parameters, e.g. $\Sigma\Delta E$ varies by a factor $\approx 5$ from $(A,D)=(0,0)$ to $(A,D)=(1,1)$.

\section*{APPENDIX B: THERMAL LOSSES}
In this paper we define $\beta=Q_{\rm p}/W_{\rm p}$ as the fraction of plastic work ($W_{\rm p}$) converted into heat ($Q_{\rm p}$). The remaining plastic work is converted into the creation of crystal defects ($E_{\rm p}$) also known as the stored energy of cold work. One sees sometimes the definition $\beta=\dot{Q}_{\rm p}/\dot{W}_{\rm p}$, where an overdot denotes a derivative with respect to time, e.g. \cite{rittel1999conversion}. The automaton in Section \ref{sec:CellAutomaton} treats failure events as instantaneous, compared to the driving (spin-down) time-scale; it averages over the highly uncertain dynamics of failure. Hence the time-integrated quantity $\beta=Q_{\rm p}/W_{\rm p}$ is more relevant in this paper. The difference between these two definitions is the difference between $(dQ_{\rm p}/dt)/(dW_{\rm p}/dt)$ versus $(\Delta Q_{\rm p}/\Delta t) / (\Delta W_{\rm p}/\Delta t)$, where $dt$ and $\Delta t$ are infinitesimal and non-infinitesimal time increments respectively. As we do not concern ourselves with the fine details of failure the non-infinitesimal definition is more appropriate for this paper. We regard the deformation as plastic to approximate crudely the global failure mode (without cracking) observed by \cite{horowitz2009breaking} and in the absence of a firm alternative.

There is more than one model of plastic deformation and more than one way to calculate $\beta$. Thermal losses have been studied in terrestrial metals, where at least one has the advantage of testing theoretical ideas against controlled experiments. The reader is referred to the thorough discussion by \cite{rosakis2000thermodynamic} for details. The stored energy of ``cold work'', $E_{\rm p}$, is attributed primarily to the creation of crystal defects. Energy balance equations consistent with the Second Law of Thermodynamics can be derived to describe the thermoplastic evolution. In brief, it is assumed that the stress, internal energy, entropy, Helmholtz free energy, heat flux and rate of plastic strain are all functions of the elastic strain, the temperature field and an internal variable called the hardening variable related to the total accumulated plastic strain. With the assumption of an adiabatic deformation and constant specific heat it is possible to derive an exact expression for $\beta$,
\begin{align}
\beta=1-\frac{\frac{dE_{\rm p}(\epsilon_{\rm p})}{d\epsilon_{\rm p}}}{\Sigma(\epsilon_{\rm p},\dot{\epsilon_{\rm p}},T)},
\end{align}
where $\Sigma$ is the \textit{stress} of deformation, $\epsilon_{\rm p}$ is the plastic strain, $\dot{\epsilon}_{\rm p}$ is the rate of plastic strain, $T$ is the temperature of the material and $E_{\rm p}(\epsilon_{\rm p})$ is the energy of cold work, which depends on quantities like the thermal softening coefficient \citep{rosakis2000thermodynamic}. There are variables beyond the scope of this appendix that specify $E_{\rm p}(\epsilon_{\rm p})$ that are typically fixed by experiment, e.g. the thermal softening coefficient.

The values of $\beta$ for terrestrial metals such as copper, steel and aluminium are approximately $0.8$. However, in addition to the factors mentioned above, the measured value can vary depending on the type of deformation (compressive, tensile or torsional), on the specific allotrope or alloy, or on the testing method. Typically $\beta$ is measured by straining a metal specimen and measuring the local temperature change using radiometric techniques, but thermocouples and calorimeters have been used too \citep{macdougall2000determination}. Naturally there is no guarantee that terrestrial metals offer a close analogy for the neutron star crust, beyond the fact that the neutron star crust is often modelled as a Coulomb crystal \citep{chamel2008physics}. In this paper, therefore, we adopt the conservative position that most of the potential energy released during plastic deformation is converted into heat, with $\beta=0.9$. Thus the mountain and gravitational wave amplitude are smaller than one would predict otherwise. The reader is encouraged to run the automaton with smaller or larger values of $\beta$, while one awaits a better understanding of the thermoplastic properties of neutron star matter. Astrophysically (as opposed to microphysically) inferred estimates of $\beta$ are rare at the time of writing. \cite{middleditch2006predicting} wrote qualitatively about post-glitch heating (see footnote 9 of the latter paper) and discussed crack formation and propagation in an appendix, but it is unclear how the latter discussion fits together with simulations of global failure \citep{horowitz2009breaking}. In a model of starquakes on Vela, \citet{bransgrove2020quake} discussed the generation of seismic waves and noted that crust-core coupling can siphon $\sim 95\%$ of the released elastic energy into the core following three rotations of the star (see text following Eq. (76) in the latter reference). This is broadly consistent with $\beta=0.9$, if one takes seismic waves as a (loose and physically distinct) proxy for dissipation during plastic deformation, as far as the global energy balance (as opposed to the local microphysics) is concerned.

\section*{APPENDIX C: Failure as a non-Poisson point process}
\begin{figure}
\includegraphics[height=5.0cm,width=8.5cm]{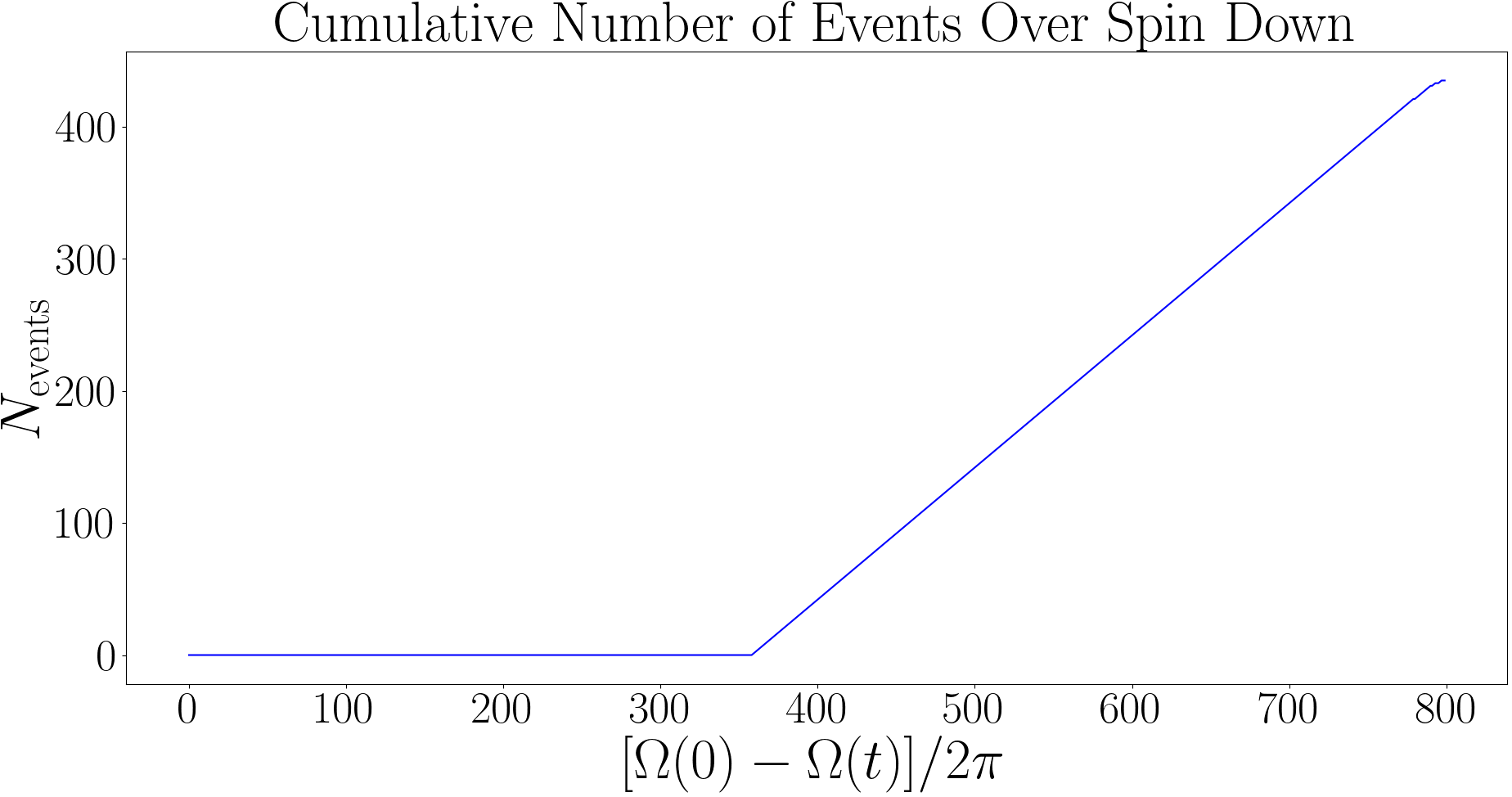}
\caption{$N(\Omega)$, the cumulative number of events occurring up to $\Omega(t)$, as a function of $\Omega(t)$ (in Hz) for $(A,D)=(0.5,0.5)$ for a single run of the simulation. The first event occurs at $\Omega(t)/2\pi= 441$ Hz, with $\Omega(0)/2\pi = 800 \, {\rm Hz}$.}
\label{fig:CumulEventsVFreq}
\end{figure}

The waiting-time PDF plotted in the top panel of Fig. \ref{fig:illustrative} depends on two pieces of physics: the mean rate at which events occur, and the statistics of the point process (e.g. Poisson) that describes the events. With respect to the mean rate, we find empirically that it is proportional to $d\Omega/dt$. Let $N(t)$ be the cumulative number of events occurring up to time $t$. Let $ N(\Omega)$ be the cumulative number of events occurring up to the corresponding angular velocity $\Omega(t)$. Examining Fig. \ref{fig:CumulEventsVFreq}, we find that $dN/d\Omega$ is a constant, after the first event occurs. Hence the chain rule implies $dN/dt = (dN/d\Omega) (d\Omega/dt) \propto d\Omega/dt$. Physically this is not surprising. For an automaton with $4\times10^4$ cells, it is probable that one or more cells are close to the failure threshold at every step in $\Omega$. A similar rate scaling proportional to $d F_{\rm d}/dt$, where $F_{\rm d}$ is the driver variable, is observed in analogous self-organized critical systems like sand piles \citep{jensen1998self}.

What about the point process? This is a harder question. The point process governing avalanches in self-organized critical systems is fundamentally unknown in general, even in the simple case of a constant driver in sand piles \citep{jensen1998self}, let alone in a neutron star where the driver decays. As an exercise, let us suppose that an inhomogeneous Poisson process is at work, with rate $\lambda(t) \propto d\Omega/dt$, i.e. $\lambda(t) = \lambda_0 (1+t/\tau)^{-3/2}$ as per the previous paragraph. Then the waiting-time PDF depends on epoch $t$ and is given by the well-known formula

\begin{eqnarray}
p(\Delta t, t) &=& \exp\left[-\int_t^{t+\Delta t} dt’ \lambda(t’)\right]\\
 &=& \exp\left\{2\lambda_0\tau\left[(1+t/\tau+\Delta t/\tau)^{-1/2}-(1+t/\tau)^{-1/2}\right]\right\}.\nonumber\\\label{eq:ExponWaitTime}
\end{eqnarray}

Eq. (\ref{eq:ExponWaitTime}) is an exponential, not a power law, in the relevant limit. Indeed, we would need $\lambda(t) = \lambda_0 (1+t/\tau)^{-1}$ to obtain a power law for a Poisson process, which would contradict Fig. \ref{fig:CumulEventsVFreq}. This implies that some non-Poisson process is at work in the neutron star problem, which is interesting albeit not surprising; self-organized critical systems contain long-range spatial correlations which translate into ``memory'' in the time domain (e.g. Omori’s law for earthquake aftershocks \citep{jensen1998self}), whereas the Poisson process is memoryless.

\section*{APPENDIX D: FREQUENCY-STEP AND CELL-NUMBER CONVERGENCE OF THE AUTOMATON}

\begin{figure}
\includegraphics[height=5.0cm,width=8.5cm]{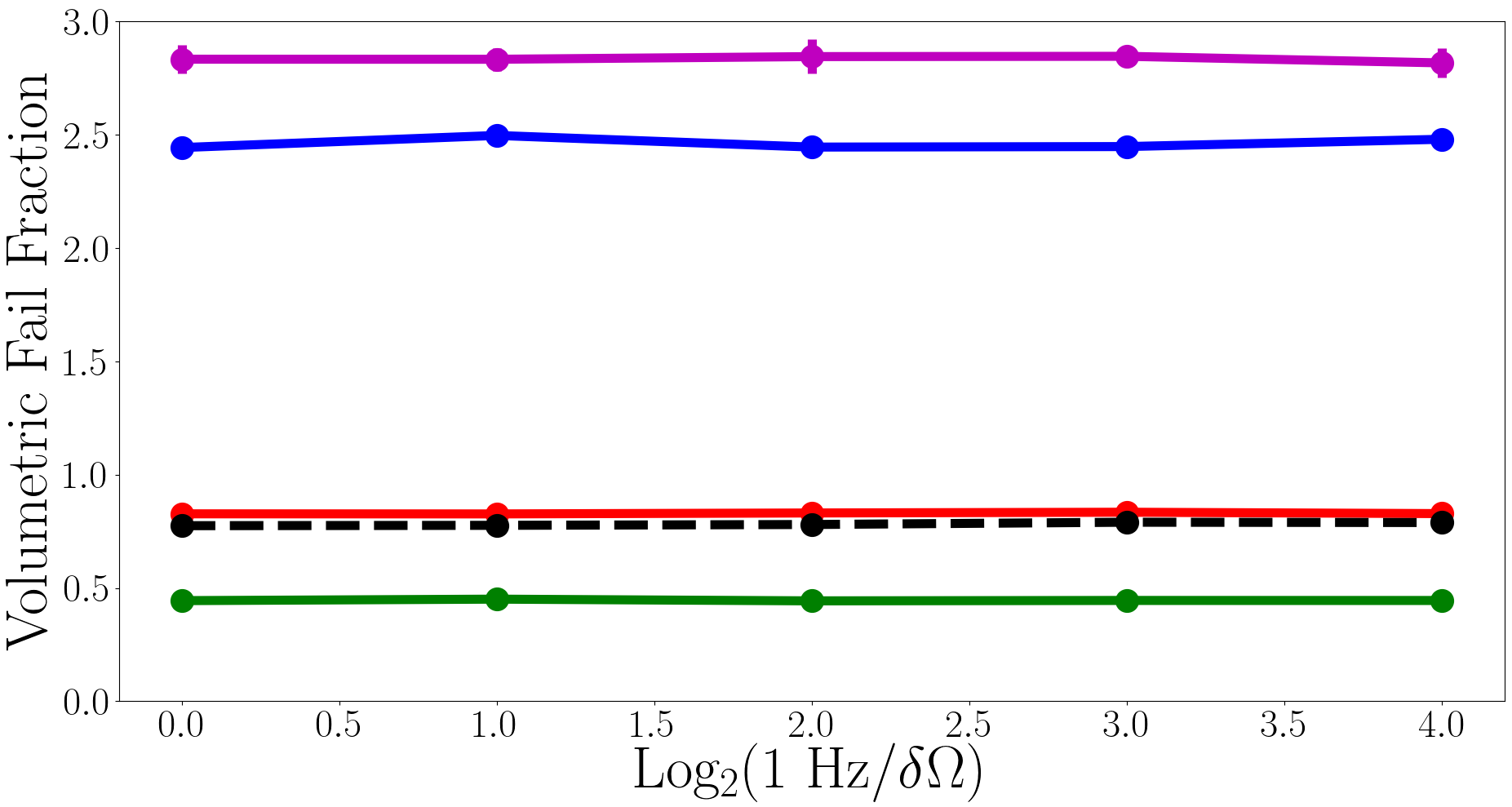}
\includegraphics[height=5.0cm,width=8.5cm]{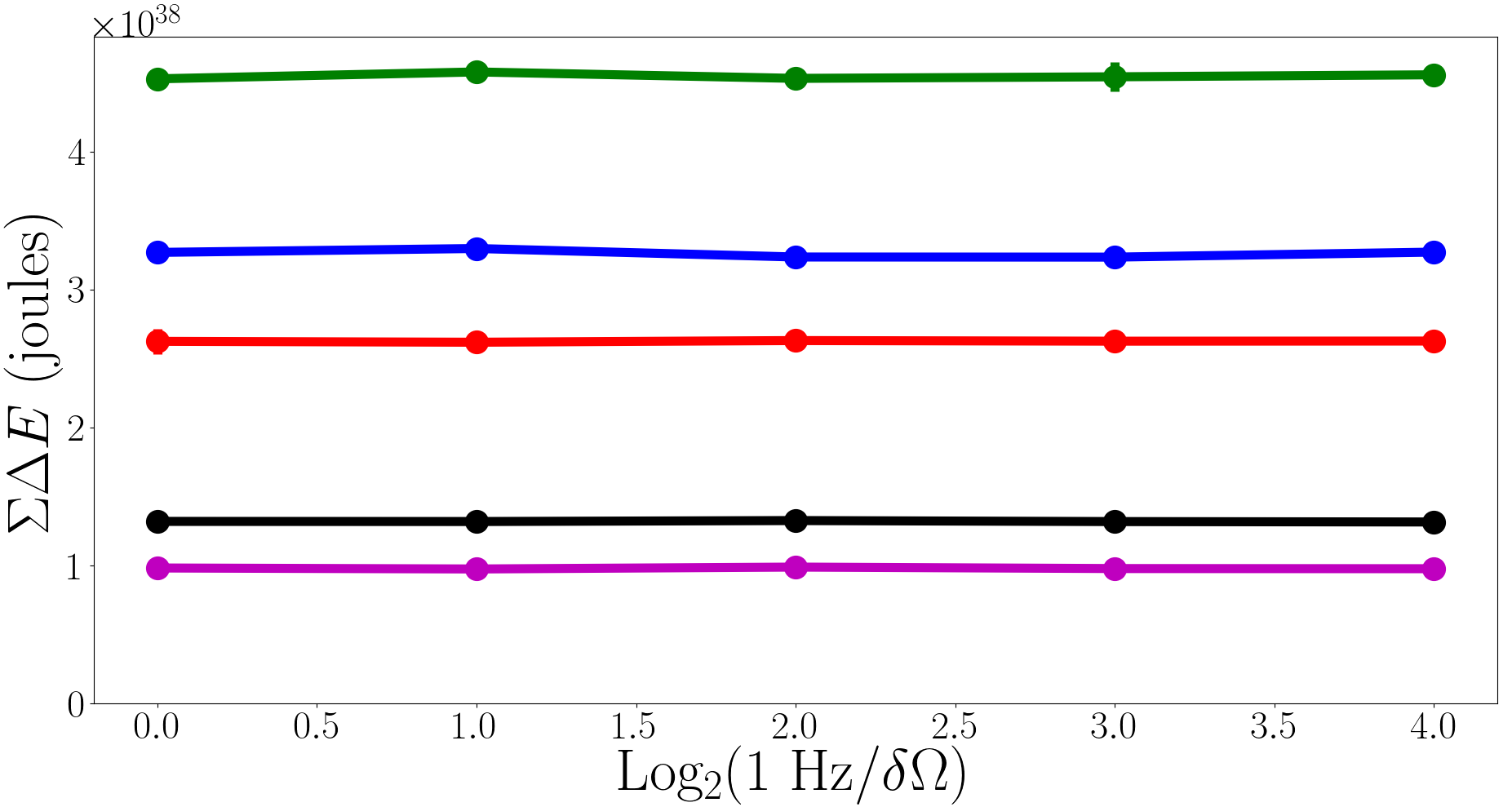}
\includegraphics[height=5.0cm,width=8.5cm]{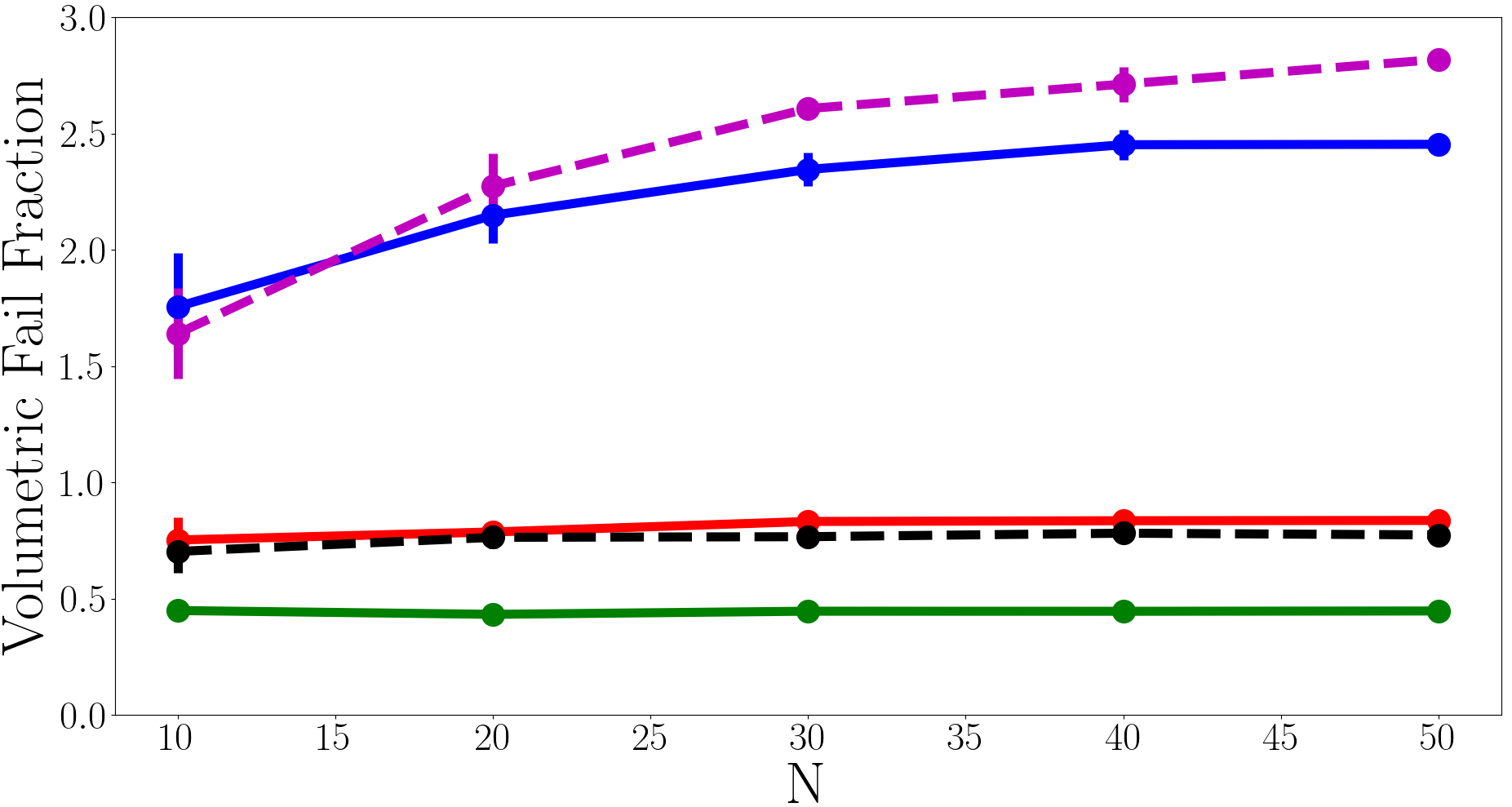}
\includegraphics[height=5.0cm,width=8.5cm]{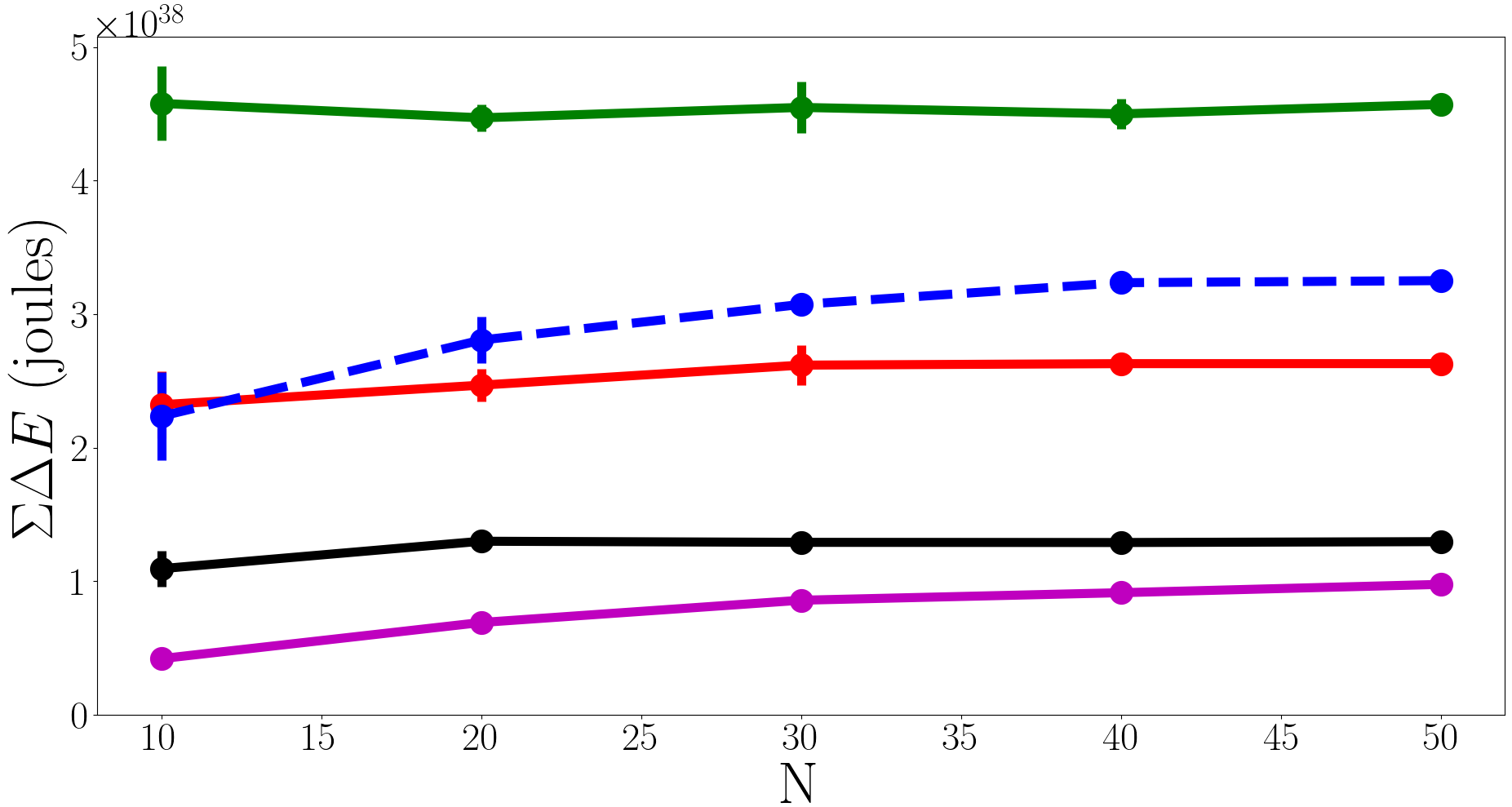}
\caption{Plotting observables against spatiotemporal resolution. The top two panels are the volume fraction of the crust that fails (first panel) and $\Sigma \Delta E$ (second panel) plotted against increasing frequency resolution. The bottom two panels are the volume fraction of the crust that fails (third panel) and $\Sigma \Delta E$ (fourth panel) plotted against increasing spatial resolution. Red corresponds to $(A,D)=(0.5,0.5)$, green to $(A,D)=(0.1, 0.1)$, blue to $(A,D)=(0.1,0.9)$, black to $(A,D)=(0.9,0.1)$, and magenta to $(A,D)=(0.9,0.9)$. Each data point corresponds to five simulations and the error bars correspond to one standard deviation. Some data are plotted with dashed lines for legibility. Unless otherwise stated one has $\delta \Omega/2\pi=0.1$Hz and $N=50$ for the data presented in this figure.}
\label{fig:Convergence}
\end{figure}
It is important to check whether or not the results of the automaton are convergent or if they are resolution-dependent, i.e. it is important to know how the number of cells ($N^2$) and the size of the frequency decrement ($\delta \Omega$) affect the final results.

In Fig. \ref{fig:Convergence} we present the fraction of crust that fails as well as $\Sigma \Delta E$ as functions of $\delta \Omega$ and $N$. In the first and second panels we see that both quantities converge as a function of $\delta \Omega$. In the third and fourth panels of Fig. \ref{fig:Convergence} we see the fraction of failed crust and $\Sigma \Delta E$ also converge with $N$.

We note that the number of events depends quadratically on $N$. The number of cells in the automaton is equal to $N^2$ and as the number of cells increases the size of each individual cell decreases, so a high-strain region of the crust contains more cells, which subsequently fail (meaning more events). However, this does not change the total heat dissipated, nor the amount of crust that fails in practice. Therefore as described in Section \ref{sec:SizeWaitDistrib} the cut-offs of the $\Delta E$ and $\Delta t/\tau$ PDFs depend on $N$ and $\delta \Omega$, with higher resolution making smaller events or shorter waiting times possible.

We find that $\delta\Omega$ does not significantly affect the behaviour of $\epsilon$ nor $h_{0}$. Both quantities converge to an accuracy of better than 25\% per cent for $N \geq 170$, recall ${\rm var}(\epsilon) \approx 0.25\langle \epsilon \rangle^2$ and ${\rm var}(h_{0}) \approx 0.25\langle h_{0} \rangle^2$.

The results presented in the body of this paper are drawn from simulations with $\delta \Omega/2\pi=1$ and $N=200$ unless otherwise specified.

\section*{APPENDIX E: MOMENT OF INERTIA CALCULATION}
In this paper the star is made of two components, the solid crust and the fluid core. In addition the crust is broken up into $N^2$ cells as part of the cellular automaton. We calculate the ellipticity and emitted wave strain of the star in Section \ref{sec:MassEllipticity} and so require its moments of inertia. We therefore need the masses and centres of mass of each crustal cell and core segment underneath each cell. Formulas for these quantities are stated for reference in this appendix.

The mass of cell $(i,j)$ is
\begin{align}
m_{i,j \rm\;crust}=&\int_{r_{i,j}}^{r_{i,j}+(R-R')}\int^{\arccos\big[\frac{\cos(\theta_{i,j})+\cos(\theta_{i+1,j})}{2}\big]}_{\arccos\big[\frac{\cos(\theta_{i-1,j})+\cos(\theta_{i,j})}{2}\big]}\nonumber\\
&\times\int_{(\phi_{i,j-1}+\phi_{i,j})/2}^{(\phi_{i,j}+\phi_{i,j+1})/2}\rho_{\rm crust}r^2\sin(\theta)d\phi d\theta dr,
\label{eq:CrustMass}
\end{align}
and the mass of the fluid segment beneath cell $(i,j)$ is 
\begin{align}
m_{i,j \rm\;core}=&\int_{0}^{r_{i,j}}\int^{\arccos\big[\frac{\cos(\theta_{i,j})+\cos(\theta_{i+1,j})}{2}\big]}_{\arccos\big[\frac{\cos(\theta_{i-1,j})+\cos(\theta_{i,j})}{2}\big]}\nonumber\\
&\times\int_{(\phi_{i,j-1}+\phi_{i,j})/2}^{(\phi_{i,j}+\phi_{i,j+1})/2}\rho_{\rm core}r^2\sin(\theta)d\phi d\theta dr.
\label{eq:CoreMass}
\end{align}

The centre of mass is defined as,
\begin{align}
\vec{C}=\frac{1}{M}\int_{V}\vec{r}\rho(\vec{r}) dV,
\label{eq:GeneralCoM}
\end{align}
where $M$ is the total mass of the object, $V$ is its volume and $\rho(\vec{r})$ is the density.
In this case the centre of mass coordinates of cell $(i,j)$ are given by
\begin{align}
\vec{C}^{i,j}_{k}=&\frac{1}{m_{i,j}}\int_{r_{i,j}}^{r_{i,j}+R-R'}
\int^{\arccos\big[\frac{\cos(\theta_{i,j})+\cos(\theta_{i+1,j})}{2}\big]}_{\arccos\big[\frac{\cos(\theta_{i-1,j})+\cos(\theta_{i,j})}{2}\big]}
\nonumber\\
&\times\int_{\frac{\phi_{i,j-1}+\phi_{i,j}}{2}}^{\frac{\phi_{i,j}+\phi_{i,j+1}}{2}}\rho_{\rm crust} k r^2\sin(\theta)d\theta d\phi dr\label{eq:CrustCoM}.
\end{align}
Likewise the center-of-mass coordinates of the underlying core segment are given by
\begin{align}
\vec{C}^{i,j}_{k}=&\frac{1}{m_{i,j}}\int_{0}^{r_{i,j}}
\int^{\arccos\big[\frac{\cos(\theta_{i,j})+\cos(\theta_{i+1,j})}{2}\big]}_{\arccos\big[\frac{\cos(\theta_{i-1,j})+\cos(\theta_{i,j})}{2}\big]}
\nonumber\\
&\times\int_{\frac{\phi_{i,j-1}+\phi_{i,j}}{2}}^{\frac{\phi_{i,j}+\phi_{i,j+1}}{2}}\rho_{\rm core} k r^2\sin(\theta)d\theta d\phi dr,\label{eq:CoreCoM}
\end{align}
with $k=x,y,z$ in Eqs. (\ref{eq:CrustCoM}) and (\ref{eq:CoreCoM}). 

Additionally we must consider the edge cases $i=0,N-1$ and $j=0,N-1$. When $i=0$ the lower terminal of integration in $\theta$ is $\theta=0$ and when $i=N-1$ the upper terminal is $\theta=\pi$ for Eqs. (\ref{eq:CrustMass}), (\ref{eq:CoreMass}), (\ref{eq:CrustCoM}), and (\ref{eq:CoreCoM}). The north and south poles act as the boundaries
of the most polar cells. Similarly when considering $j=0, N-1$, the discontinuity in $\phi$ across the positive $x$-axis must be accounted for otherwise we have $(\phi_{i,N-1}+\phi_{i,0})/2=\pi(N-1)/N$ which is inappropriate; when considering this boundary $(2\pi-\phi_{i,N-1}+\phi_{i,0})/2$ should be used instead.

Calculating the centres of mass using Eqs. (\ref{eq:CrustCoM}) and (\ref{eq:CoreCoM}) is computationally expensive. For large $N$ the centres of mass are well approximated by
\begin{align}
\vec{r}_{i,j \rm\;crust}&=\frac{3}{4}\frac{[r_{i,j}+(R-R')]^4-r_{i,j}^4}{[r_{i,j}+(R-R')]^3-r_{i,j}^3}\vec{r}_{i,j},\label{eq:SimpleCoM1}\\
\vec{r}_{i,j \rm\;core}&=\frac{3}{4}\vec{r}_{i,j}.\label{eq:SimpleCoM2}
\end{align}
The approximation is used because the computation is approximately three times faster than the exact calculation and the accuracy remains high, viz. a disagreement of $\approx 0.006\%$ for $N=200$.


\label{lastpage}
\end{document}